# Surrogate Symbols in Wireless Communications

**ZEKERIYA UYKAN, *IEEE Senior Member***
College of Engineering and Technology, American University of the Middle East, Kuwait
CORRESPONDING AUTHOR: Zekeriya Uykan (e-mail: zekeriya.uykan@aum.edu.kw).
This research received no external funding.

**ABSTRACT** Co-channel interference (CCI), caused by the inevitable aggressive channel reuse in dense deployments, remains a critical bottleneck for current and future wireless communication systems. In this paper, we address this bottleneck in an $N$-co-channel wireless communications network of single-antenna transmitters and receivers. The limiting case of this problem leads to a novel "thought experiment": If each receiver's own desired channel link is *simultaneously completely blocked*, can a Maximum Likelihood Detection at each receiver still recover its transmitted symbol? In the traditional way of thinking, the answer is no: The receiver obtains nothing but co-channel interference (CCI) plus thermal noise, and a receiver-side phase rotation cannot help, since there is no desired signal to align. Given only the channel matrix H — and no knowledge of the transmitted symbols — can the transmitters' phases be chosen so that the received signal becomes "almost" the desired symbol at every receiver? We show that the answer is yes. We first develop the underlying framework for the general case (nonzero own-channel link case), turning classical co-channel systems into a novel discrete phase-optimization problem, "Game of Phases: Surrogate Symbol Optimization", and present two algorithms, "Minimum Phased-Interference Power algorithm (MiPIP)" and "Maximum Phased-Interference Support (MaPRIS)". We then show that the MaPRIS resolves the thought experiment: it converts hostile CCI into an "almost" desired symbol even when the own desired channel link is zero, yielding the counter-intuitive result that higher CCI variance yields higher SINR. Extensive simulation results confirm the findings. The gain promised by the proposed algorithms comes on top of any other CCI reduction techniques as our framework here is based on a bottom-line baseband model.

## I. INTRODUCTION AND MOTIVATION

The demand for higher data rates, ultra-reliable connectivity and massive network capacity has pushed wireless systems into increasingly aggressive spectrum reuse scenarios in recent years. Fifth-generation (5G) and forthcoming sixth-generation (6G) cellular networks promise unprecedented data rates and spectral efficiency through technologies such as dense small-cell deployment, multiple-input multiple-output (MIMO), advanced beamforming, etc. However, despite these advancements, the co-channel interference (CCI) remains a critical unavoidable bottleneck that prevents the full realization of these performance targets. As cell densities grow and spectrum resources are reused more aggressively, the cellular systems inevitably become CCI-dominated and thus the CCI leads to a degradation of signal quality, reduced throughput, and increased latency, ultimately undermining the performance and reliability expected from modern wireless systems (e.g.[26],[27],[28],[29]). Effectively mitigating the CCI in such CCI-limited environments is essential to unlock the theoretical spectral efficiency gains envisioned for future networks.

Conventional interference management approaches—such as power control, frequency planning, and Coordinated Multi-Point (CoMP) transmission [20],[21], etc—offer partial relief but often fall short when faced with the dynamic conditions of modern cellular deployments. In this context, solving the CCI problem represents not only a technical challenge but also a key enabler that can pave the way toward achieving the full spectral efficiency potential of next-generation wireless communications. Because the traditional CCI mitigation techniques, such as power control and beamforming based methods have their limitations, the growing demand for advanced CCI management drives researchers to explore innovative techniques that leverage cutting-edge signal processing and optimization frameworks.

Power control optimization is carried out in the power domain, and most channel allocation techniques are likewise devised in the amplitude or power domain. Mitigating CCI in the phase domain, by contrast, has traditionally relied on multiple transmit and/or receive antennas — e.g., MIMO beamforming, where precise phase alignment across an antenna array improves signal-to-noise ratio (SNR) and suppresses interference [1]. Such multi-antenna phase-domain techniques are orthogonal to, and outside the scope of, this paper. Here we instead ask a more basic question: In an $N$-co-channel network of single-antenna transmitters and receivers — relying only on a single discrete phase assigned to each tx and rx, with no antenna-array gain — is symbol recovery possible at all when every receiver's own desired channel link is *simultaneously and completely blocked*?

It is worth emphasizing that this ***single-antenna***, phase-only restriction is not a competing alternative to power control, beamforming, CoMP, interference cancelation, etc, but a layer that operates independently of them. Power control, CoMP, beamforming, interference reduction/cancelation and other radio resource management algorithms operate at a different layer and shape the channel matrix **H** itself. On the other hand, the channel matrix **H** in our paper refers to the *effective* channel matrix already obtained after any such algorithm has been applied, and not the absolute physical channel. Because MiPIP and MaPRIS reshape only the residual phase relationships in this already-formed channel and baseband model, the gains reported in this paper come on top of, rather than in place of, any other CCI mitigation technique the system may already employ.

Inspired by the concept called shadow-cuts in [2], we, in this paper, reformulate/turn classical co-channel wireless communication systems into a novel phase-optimization framework, where each transmitter and receiver is associated with a discrete phase. The concept of phase plays a fundamental role in various scientific and engineering disciplines, including physics, signal and image processing, optics, control systems, neuroscience, and quantum mechanics [4],[5]. Phase-based filtering enhances edge detection and feature extraction in image processing; phase coherence enables ultra-precise interferometric measurement in optics; and phase synchronization plays a vital role in information processing in physiological neural networks.

In wireless communications specifically, phase interactions are critical in electromagnetic wave propagation: multiple waves can cancel each other out when in opposite phases, or reinforce each other when in phase. This behavior underlies beamforming, interference cancellation, and signal synchronization techniques. In this paper, we show that modeling and exploiting phase-dependent interactions among co-channel transmitters and receivers is itself sufficient to optimize wireless system performance — without requiring antenna arrays.

This raises a question that, at first glance, seems hopeless to answer in the affirmative in our “though experiment”. Consider a receiver whose own desired channel link is completely blocked, i.e., the desired channel coefficient is zero. In the traditional way of thinking, no receiver-side signal processing can recover the transmitted symbol in this case: The receiver's input consists of nothing but CCI and thermal noise, and rotating a zero desired signal cannot help. Now consider an N-co-channel-interference network in which this own-channel blockage occurs simultaneously at every receiver. Given only the channel matrix H — and no knowledge of the transmitted symbols at all — can the transmitters' phases nonetheless be chosen so that, at every receiver, the received signal becomes "almost" the desired symbol at the expense of an extra receiver phase info before the MLD? We show in this paper that the answer is yes. This "thought experiment" is not just a mathematical curiosity. It shows, in its most extreme form, the central idea of this paper: CCI does not have to be treated as a hostile quantity to be suppressed. Instead, phase optimization can reshape it into a stand-in for the desired signal itself.

To resolve this thought experiment, and more generally to address the co-channel wireless system in which the desired channel link is nonzero, we introduce a novel "surrogate symbol" transmission framework, wherein a proxy symbol is selected and transmitted instead of the intended (optimized) desired symbol. Our investigations, under some assumptions, yield the following novel results:

### A. CONTRIBUTION OF THIS PAPER

- We pose and resolve a "thought experiment": even when every receiver's own desired channel link is completely blocked in an $N$-co-channel network — a case in which classical receivers have, in principle, no way to recover the transmitted symbol — we show that the desired symbol can still be recovered with high probability, by converting the co-channel interference itself into an "almost" desired symbol using only the channel matrix H and a receiver phase index.
- More generally, inspired by the concept of shadow-cuts in [2], we reformulate classical co-channel wireless systems into a novel phase-optimized framework, where each tx and rx is associated with a discrete phase state. This leads to a novel optimization problem, "Game of Phases: Surrogate Symbol Optimization": given a channel gain matrix and a discrete phase set, we seek the optimal surrogate symbols ("proxies") that minimize average Bit Error Rate (BER) / maximize average Signal-to-Interference+Noise Ratio (SINR) of the overall wireless system.
- We present two novel phase-optimization algorithms for solving this problem: Minimum Phased-Interference Power (MiPIP) and Maximum Phased-Interference Support (MaPRIS).

- We show that MiPIP reduces BER by minimizing the variance of the phased interference.
- We show that MaPRIS converts "hostile" CCI into a "friendly signal" close to the desired symbol — a mechanism we call "interference support" — yielding the counter-intuitive result that higher CCI variance produces higher SINR and thus higher data rate.
- We introduce the "effective angle," a novel concept measuring how "friendly" the phased-and-rotated CCI is to the desired symbol.
- Extensive simulation results, for both the general ($\boldsymbol{h}_{ii} \neq 0$) and fully-blocked ($\boldsymbol{h}_{ii} = 0$) cases, confirm all findings. The best results are obtained by the MaPRIS by far in both cases. For the fully-blocked ($\boldsymbol{h}_{ii} = 0$) case, the 2nd best results are obtained by the Complex Hopfield Neural Network and SIGA [2] based phase optimization algorithms.
- For tractability, our analysis assumes an idealized centralized-CSI ("genie") setting for establishing performance limits. The reported results are therefore an upper bound on the performance of practical systems. We discuss the practical signaling implications of this assumption, and directions for relaxing it in appendix A.

This article is arranged as follows: the traditional $N$ co-channel case ($\boldsymbol{h}_{ii} \neq 0$) is given in Section II. . We present the surrogate symbol concept and the proposed surrogate-symbol-based wireless system in Section III. . The novel algorithms MiPIP and MaPRIS are presented in Section IV. . We then resolve the thought experiment by examining the fully-blocked own-channel case ($\boldsymbol{h}_{ii} = 0$) as a special case of the proposed framework in Section V. Computer simulation results for both cases are given in Section VI. , followed by conclusions in Section VII. .

For the reader's convenience, we list the abbreviations used in this paper in Table 1:

**Table 1.** Abbreviations in this paper.

| Name | Phase algorithm |
|---|---|
| CCI | Co-Channel Interference |
| BER | Bit Error Rate |
| SER | Symbol Error Rate |
| SINR | Signal to Interference+Noise Ratio |
| Tx | Transmitter |
| Rx | Receiver |
| MLD | Maximum Likelihood Detection |
| MiPIP | Minimum Phased-Interference Power algorithm |
| MaPRIS | Maximum Phased-Interference-Support algorithm |
| QAM | Quadrature Amplitude Modulation |
| CoMP | Coordinated Multi-Point |

## II. TRADITIONAL CO-CHANNEL TRANSMITTERS CASE

In conventional wireless communication systems, the process of determining the optimum transmit symbol follows a structured sequence of operations, including source coding, channel coding, modulation, and symbol selection. A brief overview of these key steps is as follows: Source Coding compresses the original data (e.g., voice, video, or text) to remove redundancy, reducing the required bandwidth. Channel coding introduces redundancy through error-correcting codes, enhancing the robustness of transmission against noise and interference. Then the modulation maps the encoded data onto signal waveforms such as Quadrature Amplitude Modulation (QAM). The transmitter selects the symbol that maximizes the likelihood of correct reception based on the channel conditions, employing techniques like Maximum Likelihood Detection (MLD) to minimize the bit error probability. The optimized symbol is finally transmitted over the channel.

In this paper, we use the complex baseband equivalent model, which is a fundamental model for simulating wireless communication systems. The channel including the effects of the transmit chain, the wireless environment, and the receive chain is modeled by a complex multiplicative distortion composed of a magnitude response and a phase response. The channel between the transmitter $i$ and the receiver $j$ is denoted by $h_{ij}$. At a given time, the optimized signal $s_i$ is sent from the transmitter $i$.

### A. TRADITIONAL 1 TX-RX PAIR CASE

The classical 1 Tx-Rx wireless system is shown in FIGURE 1. Noise and interference are added at the receiver. The resulting received baseband signal is

$$\boldsymbol{r}_1 = \boldsymbol{s}_1\boldsymbol{h}_{11} + \boldsymbol{n}_1 \tag{1}$$

where $\boldsymbol{h}_{11}$ is the baseband channel coefficient and is represented by

$$\boldsymbol{h}_{11} = \alpha_{11}e^{j\theta_{11}} \text{ and } \boldsymbol{s}_1 = E_s e^{j\phi_1} \tag{2}$$

where $E_s$ represents the energy of the symbol. In this paper, for the sake of clarity, we use the notation of Alamouti in [1]. Assuming $n_1$ is Gaussian distributed, the maximum likelihood decision rule at the receiver for the received signal $r_1$ is to choose signal $\boldsymbol{s_i}$ if and only if

$$d^2(\boldsymbol{r}_1,\ \boldsymbol{s}_i\boldsymbol{h}_{11}) \leq d^2(\boldsymbol{r}_1,\ \boldsymbol{s}_k\boldsymbol{h}_{11}) \ , \quad \forall i \neq k \tag{3}$$

where $d^2(\boldsymbol{x},\boldsymbol{y})$ is the squared Euclidean distance between signals $\boldsymbol{x}$ and $\boldsymbol{y}$ calculated by the following expression

$$d^2(\boldsymbol{x},\boldsymbol{y}) \leq (\boldsymbol{x}-\boldsymbol{y})(\boldsymbol{x}^*-\boldsymbol{y}^*) \tag{4}$$

It is well known that the optimum weight at the receiver is the conjugate of the channel, which is estimated by the channel estimator. So, the input to the MLD is

$$\begin{aligned}\tilde{\boldsymbol{s}}_1 &= \frac{\boldsymbol{h}_{11}^*}{\alpha_{11}}(\boldsymbol{s}_1\boldsymbol{h}_{11} + \boldsymbol{n}_1) \\ &= \alpha_{11}\boldsymbol{s}_1 + e^{-j\theta_{11}}\boldsymbol{n}_1\end{aligned} \tag{5}$$

where $e^{-j\theta_{11}} = \frac{\boldsymbol{h}_{11}^*}{\alpha_{11}}$.

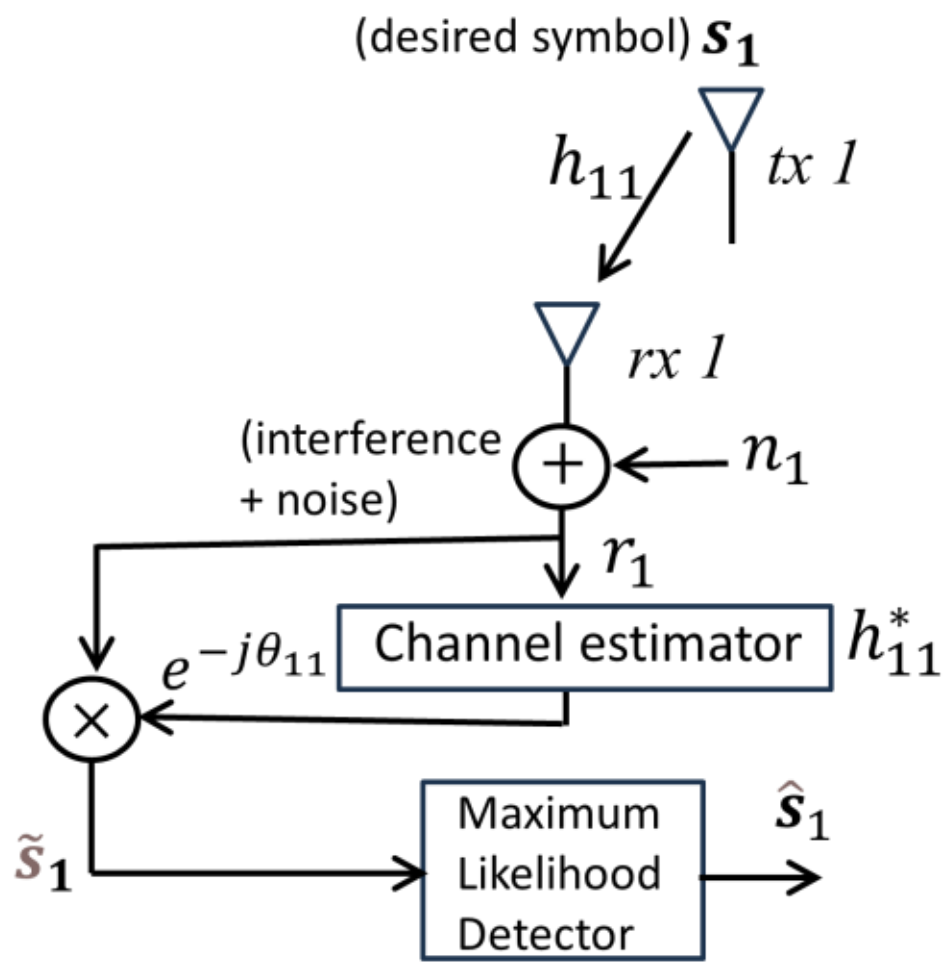

**FIGURE 1 Traditional 1-antenna tx-rx case.**

For PSK signals (equal energy constellations)

$$E_s = |s_k|^2 \quad , \quad \forall k \tag{6}$$

where $E_s$ is the energy of the signal. Therefore, using (3), (4) and (5), the MLD decision rule for PSK signals can be simplified as follows:

Choose $\hat{s}_1 = s_i$ if and only if

$$d^2(\tilde{s}_1, s_i) \leq d^2(\tilde{s}_1, s_k) \quad , \quad \forall i \neq k \tag{7}$$

So, the MLD gives $\hat{s}_1 = s_i$, which is a maximum likelihood estimate of the desired symbol $s_1$.

## B. TRADITIONAL N (CO-CHANNEL) RX-TX PAIRS CASE

Traditional $M$ co-channel tx-rx pairs case is shown in FIGURE 2. Each tx has one antenna and each rx has only one antenna.

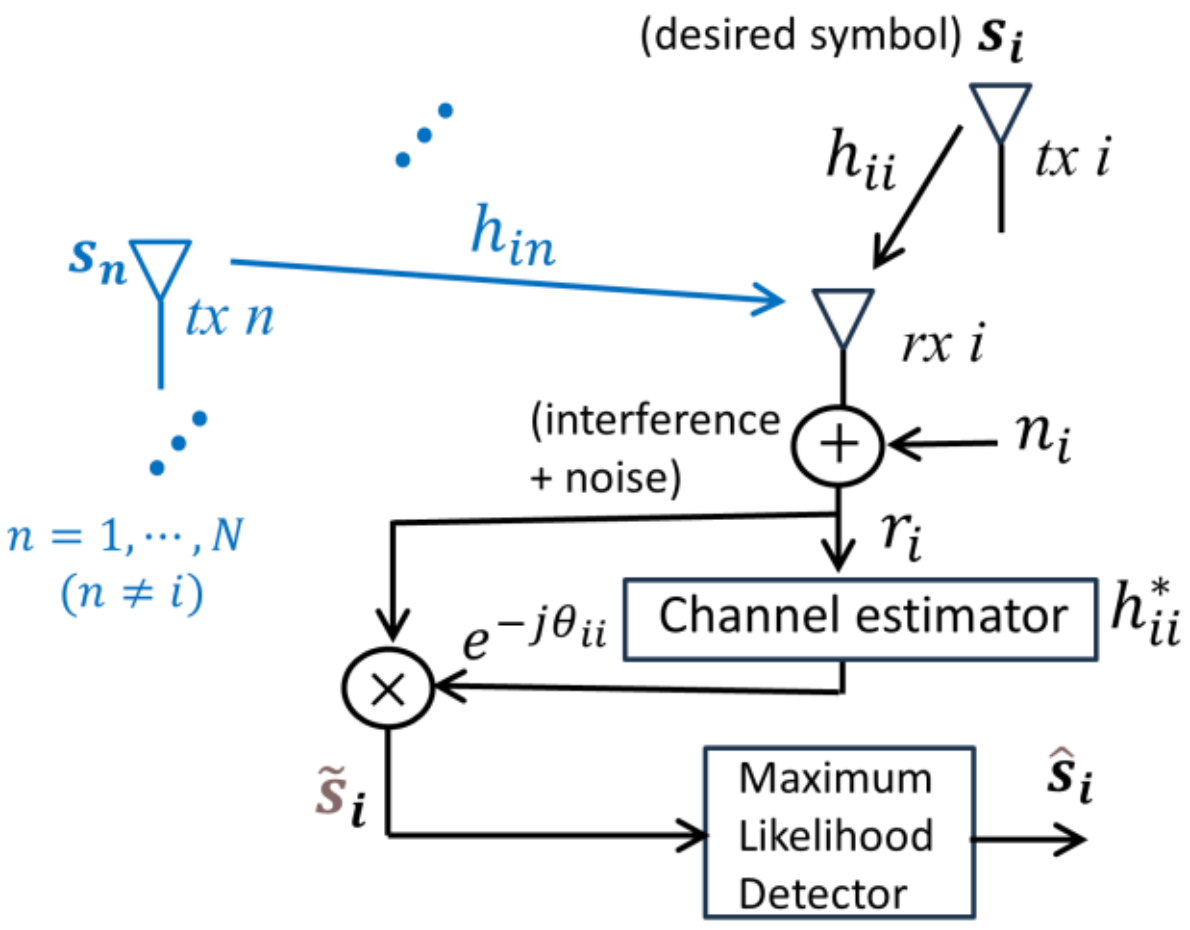

**FIGURE 2 Traditional communication system (*N* co-channel one-antenna tx-rx pairs case).**

Let's represent the baseband channel coefficient between tx $n$ and rx $i$ as

$$h_{in} = \alpha_{in} e^{j\theta_{in}} \tag{8}$$

Adding the noise and co-channel interference (CCI), the resulting received baseband signal for rx $i$ in FIGURE 2 is as follows:

$$r_i = s_i h_{ii} + I_i + n_i \tag{9}$$

where $I_i$ is the CCI for rx $i$,

$$I_i = \sum_{\substack{n=1 \\ (n \neq i)}}^{N} s_n h_{in} \tag{10}$$

$$I_i = \sum_{\substack{n=1 \\ (n \neq i)}}^{N} s_n \, \alpha_{in} e^{j\theta_{in}}$$

and noise $n_i$ is Gaussian distributed. From eqs.(8)-(9), and **FIGURE 2**, input to the MLD is

$$\tilde{s}_i = r_i \frac{h_{ii}^*}{\alpha_{ii}} \tag{11}$$

$$\tilde{s}_i = \alpha_{ii} s_i + e^{-j\theta_{ii}}(I_i + n_i)$$

Following the steps in eqs.(3) to (5) yields similar MLD rule as in previous section for the PSK signals (equal energy constellations):

Choose $\hat{s}_i = s_m$ if and only if

$$d^2(\tilde{s}_i, s_m) \leq d^2(\tilde{s}_i, s_k) \quad , \quad \forall m \neq k \tag{12}$$

The MLD gives $\hat{s}_i = s_m$, which is a maximum likelihood estimate of the desired symbol $s_i$.

# III. SURROGATE SYMBOLS IN WIRELESS COMMUNICATIONS AND GAME OF PHASES

In this section, we present our prosed surrogate symbol based wireless communication system to mitigate the CCI where each transmitter has only one antenna.

Recently, the author in [2] investigated directed graphs with complex-valued edges where each node is assigned a discrete phase. Directed graphs allow two edges to join the same vertex, but they must go in opposite directions. This directed-graph representation of the co-channel system follows standard graph-theoretic conventions for weighted directed graphs [9], where nodes correspond to transmitter–receiver pairs and directed edges correspond to the (asymmetric) channel coefficients between them. In [2], the "*sum of phased-inbound-edges*" corresponds to the CCI $I_i$ in (10). The "*inter-*cluster *phased*-edges" in [2] corresponds to channel-specific CCI in multi-channel scenario. The sum of those *inter*-cluster *phased*-edges is called as "shadow cut". Being inspired by the findings in [2], we define the following:

**Definition**: **Phase Set:** The set of $K$ discrete phases for the transmitter-receiver pairs is called **phase set** represented by $P = \{e^{-i\varphi_1}, e^{-i\varphi_2}, \cdots, e^{-i\varphi_K}\}$. The elements of the phase set are obtained by evenly dividing the complex unit circle into $K$ regions and the centres of these $K$ regions give the phase set $\{e^{-i\varphi_k}\}_{k=1}^{K}$.

This discrete-state formulation is conceptually related to classical discrete-state models in statistical physics, where each site (or spin) of a lattice is restricted to one of a finite set of states — most notably the two-state Ising model [6] and its K-state generalization, the Potts model [8], both of which have been extensively analyzed in terms of order-disorder phase transitions [7]. In our setting, each transmitter–receiver pair plays the role of such a discrete-state site, with the phase set *P* taking the place of the spin/state alphabet.

**Definition**: **Transmit phase and Receive phase**: The phase of transmit *i* at a specific time is represented by $\boldsymbol{\varphi}_{p(i)}^{tx}$ (i.e. surrogate symbol's phase) where $i \in \{1,2,\cdots,N\}$ and $p(i) \in \{1,2,\cdots,K\}$. Corresponding receiver's phase is represented as $(-\boldsymbol{\varphi}_{p(i)}^{rx})$.

Inspired by the shadow-cut concept in [2], we propose to assign each transmitter and each receiver one of these phases at a given channel use.

**Definition**: ***Surrogate Symbol*:** For transmitter *n,* we define *surrogate symbol* as shifted version of the (optimized) desired symbol as follows:

$$\boldsymbol{s}_i^{sur} = c_i e^{j\boldsymbol{\varphi}_{p(i)}^{tx}} \quad (13)$$

where $c_i$ is the energy of the surrogate symbol and $\boldsymbol{\varphi}_{p(i)}^{tx}$ is the phase of the surrogate symbol. The relation between the tx phase, rx phase and the desired symbol $\mathbf{s}_i$ is given by

$$\boldsymbol{s}_i^{sur} = c_i e^{j\varphi_{p(i)}^{tx}} = e^{j\varphi_{p(i)}^{rx}} \boldsymbol{s}_i \quad (14)$$

Without loss of generality and for the sake of brevity, in this paper we assume $c_i = 1$.

### *A. PROPOSED 1 TX-RX PAIR CASE*

As in [2], we propose to assign a specific discrete phase to each tx and each rx. Surrogate symbol's phase of tx *i* is $\boldsymbol{\varphi}_{p(i)}^{tx}$ and corresponding receiver's phase is $(-\boldsymbol{\varphi}_{p(i)}^{rx})$. The proposed one-antenna one tx-rx case is shown in **FIGURE 3**.

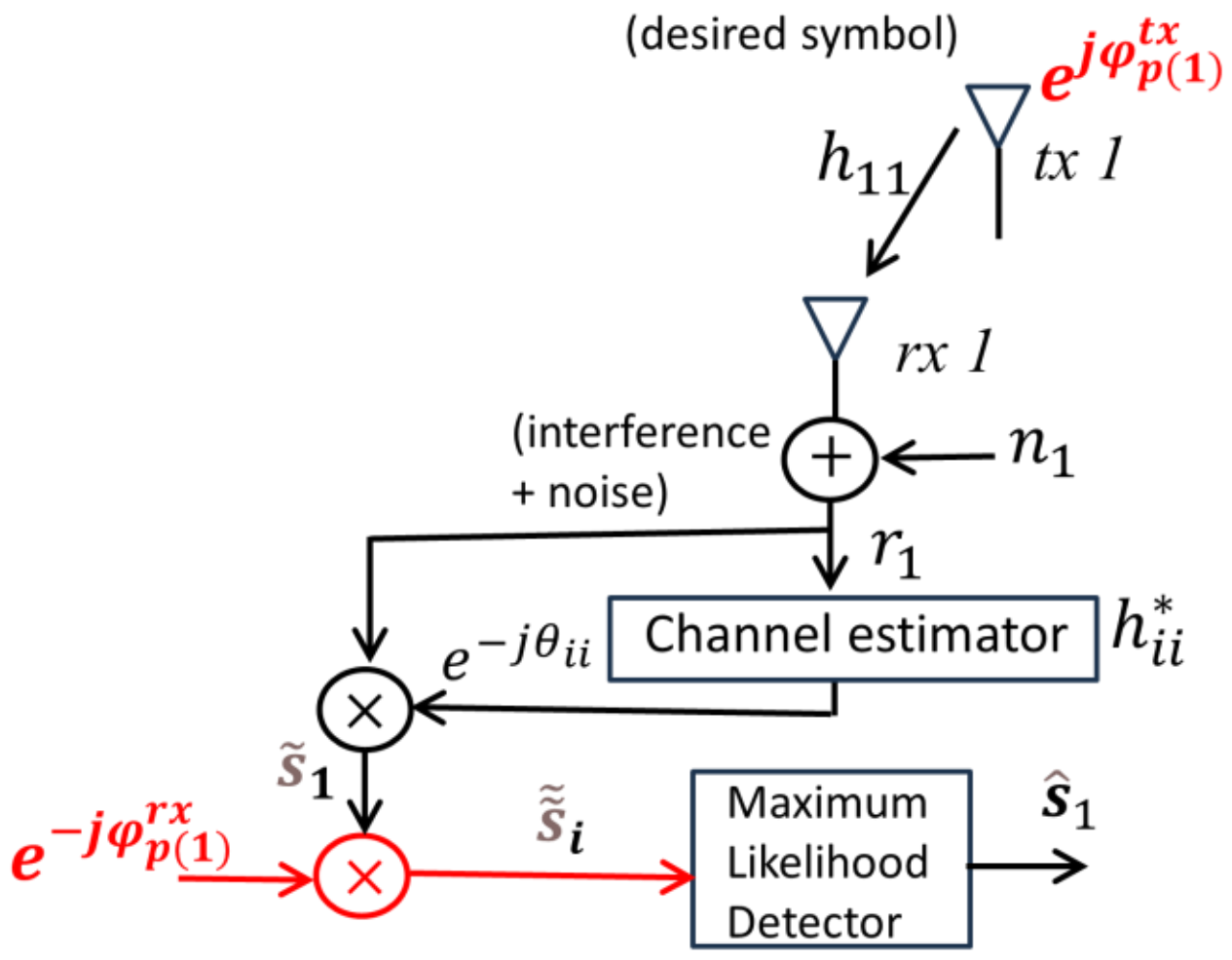

**FIGURE 3 Proposed 1 Tx-Rx pair case.**

Adding the noiseat the receiver, the resulting received baseband signal is

$$\boldsymbol{r}_1 = \boldsymbol{s}_1^{sur}\boldsymbol{h}_{11} + \boldsymbol{n}_1 = c_i e^{j\varphi_{p(1)}^{tx}}\boldsymbol{h}_{11} + \boldsymbol{n}_1 \quad (15)$$

where $\boldsymbol{h}_{11}$ is the baseband channel coefficient and is represented by (2). Using (15), after the channel estimation in **FIGURE 3**, we have

$$\tilde{\boldsymbol{s}}_1 = \frac{\boldsymbol{h}_{11}^*}{\alpha_{11}}\boldsymbol{r}_1 \quad (16)$$

$$\tilde{\boldsymbol{s}}_1 = e^{-j\theta_{11}}\left(e^{j\varphi_{p(1)}^{tx}}\boldsymbol{h}_{11} + \boldsymbol{n}_1\right)$$

where $e^{-j\theta_{11}} = \frac{\boldsymbol{h}_{11}^*}{\alpha_{11}}$. According to the proposed scheme in FIGURE **3**, we multiply the signal $\tilde{\boldsymbol{s}}_1$ by $e^{-j\varphi_{p(1)}^{rx}}$ just before the MLD. Using eqs.(13)-(16), we obtain the input signal to the MLD as follows:

$$\tilde{\tilde{\boldsymbol{s}}}_1 = e^{-j\varphi_{p(1)}^{rx}}\,\tilde{\boldsymbol{s}}_1 = \alpha_{11}\boldsymbol{s}_1 + e^{-j\left(\varphi_{p(1)}^{rx}+\theta_{11}\right)}\boldsymbol{n}_1 \quad (17)$$

where $\boldsymbol{h}_{11}^* = \alpha_{11}e^{-j\theta_{11}}$ from (8) and $\boldsymbol{s}_1$ is the desired symbol to be transmitted. Because $\boldsymbol{n}_1$ is Gaussian distributed in (15), following the steps in (3) to (5) yields similar maximum likelihood decision rule in section II. A. :

Choose $\hat{\boldsymbol{s}}_1 = \boldsymbol{s}_m$ if and only if

$$d^2(\tilde{\tilde{\boldsymbol{s}}}_1, \boldsymbol{s}_m) \le d^2(\tilde{\tilde{\boldsymbol{s}}}_1, \boldsymbol{s}_k), \quad \forall m \ne k \quad (18)$$

The signal $\tilde{\tilde{\boldsymbol{s}}}_1$ is then constructed as shown in **FIGURE 3**, so that the MLD produces $\hat{\boldsymbol{s}}_1 = \boldsymbol{s}_m$, which is the maximum likelihood estimate of $\boldsymbol{s}_1$.

***Corollary 1:*** The Surrogate Symbol based proposed wireless system of 1 tx-rx pair in FIGURE **3** gives the same Bit Error Rate (BER) performance as the traditional wireless system in **FIGURE 1** does.

***Proof:*** The proof of the corollary is straightforward and follows from the fact that only rotating the gaussian random noise does not change its variance. So, the proposed wireless system in FIGURE **3** yields the same bit error rate performance as in the system in **FIGURE 1**.

■

### *B. PROPOSED N (CO-CHANNEL) RX-TX PAIRS CASE*

Introducing the surrogate symbol concept and correspondingly assigning each tx and each rx a discrete phase as in [2], we propose the *N* (co-channel) tx-rx pairs wireless systems in FIGURE **4**.

**Proposition 1:** In both the traditional co-channel wireless system in **FIGURE 2** and in the proposed surrogate-symbol based wireless system in FIGURE **4**, the input to the MLD is in the following form:

$$\boldsymbol{s}_i^{MLD} = \alpha_{ii}\boldsymbol{s}_i + \boldsymbol{I}_i^{rot} + \boldsymbol{n}_i^{rot} \quad (19)$$

where $\boldsymbol{s}_i$ is the desired symbol to be transmitted and $\boldsymbol{I}_i^{rot}$ and $\boldsymbol{n}_i^{rot}$ represents a rotated CCI signal and a rotated noise signal, respectively. To be precise, in the traditional wireless system in **FIGURE 2**:

$$\boldsymbol{I}_i^{rot} = e^{-j\theta_{ii}}\boldsymbol{I}_i \text{ and } \boldsymbol{n}_i^{rot} = e^{-j\theta_{ii}}\boldsymbol{n}_i \quad (20)$$

and in the proposed surrogate-symbol based wireless system in FIGURE **4**:

$$\boldsymbol{I}_i^{rot} = c_i e^{-j\left(\varphi_{p(i)}^{rx}+\theta_{ii}\right)} \boldsymbol{I}_i^{sur} \text{ and } \quad (21)$$

$$\boldsymbol{n}_i^{rot} = e^{-j\left(\varphi_{p(i)}^{rx}+\theta_{ii}\right)} \boldsymbol{n}_i$$

where $\boldsymbol{I}_i^{sur} = \sum_{\substack{n=1 \\ (n\neq i)}}^{N} \boldsymbol{s}_n^{sur} \boldsymbol{h}_{in}$.

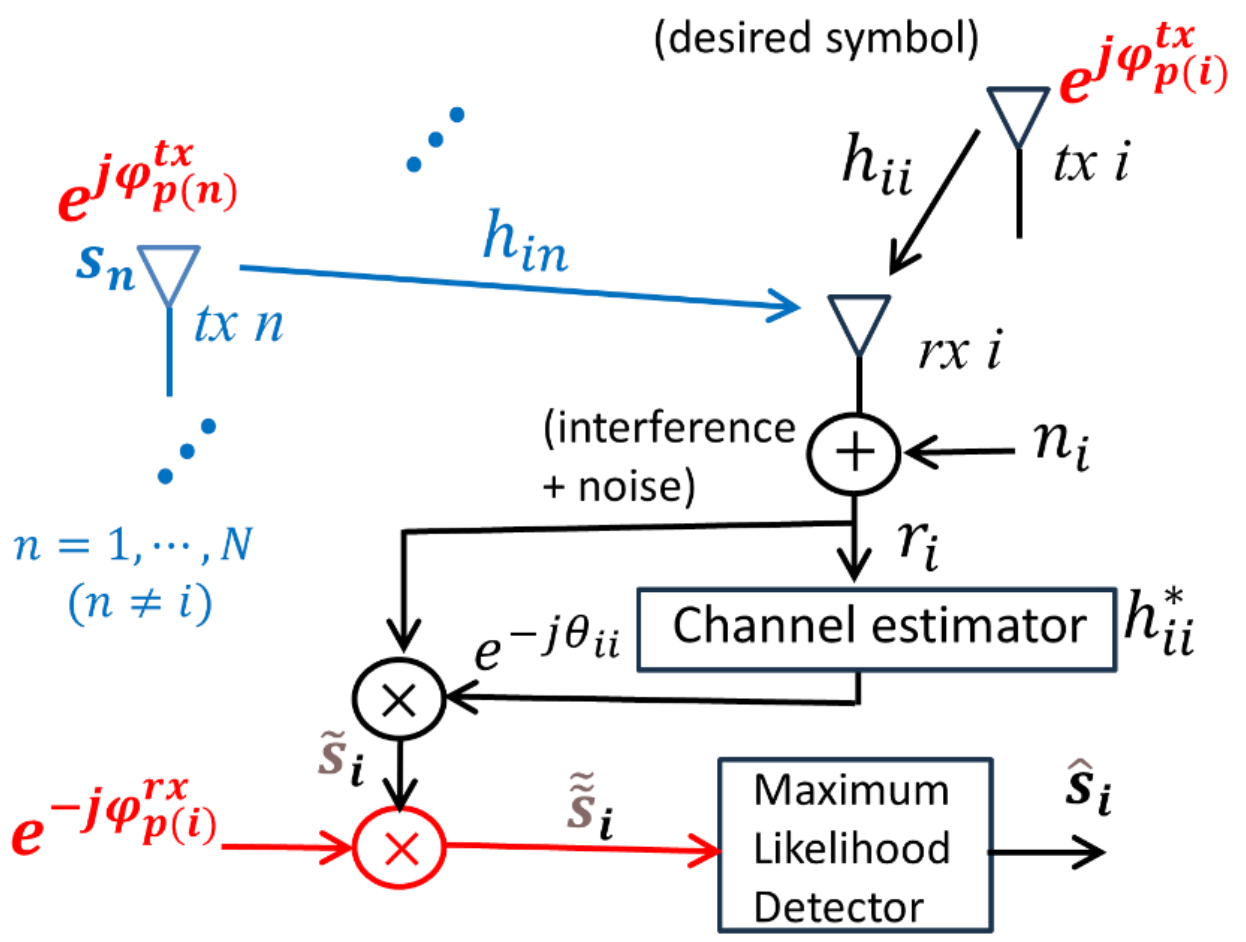

**FIGURE 4 Proposed surrogate-symbol based communication system (*N* co-channel one-antenna tx-rx pairs case).**

***Proof:*** Consider the proposed surrogate-based wireless system in **FIGURE 4**: Adding the noise and CCI at the receiver $i$, the resulting received baseband signal is

$$\boldsymbol{r}_i = \boldsymbol{s}_i^{sur} \boldsymbol{h}_{ii} + \boldsymbol{I}_i^{sur} + \boldsymbol{n}_i \quad (22)$$

where $\boldsymbol{h}_{ii}$ is the baseband channel coefficient represented by (2) and

$$\boldsymbol{I}_i^{sur} = \sum_{\substack{n=1 \\ (n\neq i)}}^{N} c_i e^{j\varphi_{p(n)}^{tx}} \boldsymbol{h}_{in} \quad (23)$$

$$i = 1,2,\cdots,N$$

where $\boldsymbol{s}_i^{sur} = c_i e^{j\varphi_{p(i)}^{tx}}$. Following the steps in (3) to (5) yield similar maximum likelihood decision rule as in previous section: After the channel estimation

$$\tilde{\boldsymbol{s}}_i = \frac{\boldsymbol{h}_{ii}^*}{\alpha_{ii}} \boldsymbol{r}_i \quad (24)$$

$$\tilde{\boldsymbol{s}}_i = e^{-j\theta_{ii}} (\boldsymbol{s}_i^{sur} \boldsymbol{h}_{ii} + \boldsymbol{I}_i^{sur} + \boldsymbol{n}_i)$$

where $e^{-j\theta_{ii}} = \frac{\boldsymbol{h}_{ii}^*}{\alpha_{ii}}$. According to the proposed scheme in **FIGURE 4**, we multiply the signal $\tilde{\boldsymbol{s}}_i$ by the rx-phase $e^{-j\varphi_{p(i)}^{rx}}$ just before the MLD. Using (13)-(14), (24), we finally obtain

$$\tilde{\tilde{\boldsymbol{s}}}_i = e^{-j\varphi_{p(i)}^{rx}} \tilde{\boldsymbol{s}}_i = \alpha_{ii} \boldsymbol{s}_i + e^{-j\left(\varphi_{p(i)}^{rx}+\theta_{ii}\right)} \boldsymbol{I}_i^{sur} + e^{-j\left(\varphi_{p(i)}^{rx}+\theta_{ii}\right)} \boldsymbol{n}_i \quad (25)$$

]where $\boldsymbol{h}_{ii}^* = \alpha_{ii} e^{-j\theta_{ii}}$ and $\boldsymbol{s}_i$ is the desired symbol to be transmitted.

Eqs.(19)-(20) are directly obtained from (11) which is for the traditional wireless system in **FIGURE 2** . On the other hand, eq.(25), obtained from the surrogate-symbol based wireless system in FIGURE **4**, yields (19) and (21), which completes the proof. ■

Using Proposition 1, we conclude that the MLD rule in section II. A. for the traditional wireless system also applies to the proposed surrogate-symbol based wireless system in FIGURE **4**: Following the steps (3) to (5) yields the following rule: Choose $\boldsymbol{s}_m$ if and only if

$$d^2(\tilde{\tilde{\boldsymbol{s}}}_i, \boldsymbol{s}_m) \leq d^2(\tilde{\tilde{\boldsymbol{s}}}_i, \boldsymbol{s}_k), \quad \forall m \neq k \quad (26)$$

The signal $\tilde{\tilde{\boldsymbol{s}}}_i$ is then constructed as shown in **FIGURE 4**, so that the MLD produces $\hat{\boldsymbol{s}}_i = \boldsymbol{s}_m$, which is the maximum likelihood estimate of the desired symbol $\boldsymbol{s}_i$.

As mentioned above, the proposed wireless system is inspired by the Shadow-Cut concept in [2]. In order to comprehend the similarities and differences between the two concepts, Table 2 below presents a comparison.

**Table 2** Comparison of the shadow-cut concept in [2] and the proposed wireless system in **FIGURE** 4.

| **Shadow-Cut paper [2]** | **Proposed *M* co-channel wireless systems** |
|---|---|
| $w_{nm}$: complex edge from node $m$ to node $n$ in the directed graph in [2] | $h_{nm}$: baseband channel coefficient between tx $m$ and rx $n$ in the wireless system |
| Matrix is random but Complex Hermitian ($w_{jk}$=$w_{jk}^*$) | Matrix is random complex matrix, which is not Hermitian. |
| sum of *phased*-inbound-edges for vertex *n in [2]* $\sum_{\substack{n=1 \\ (n\neq i)}}^{N} w_{in} e^{j\varphi_{p(i)}}$ | sum of *phased*-CCI for rx *i*: *(eq.(23) this paper)* $\sum_{\substack{n=1 \\ (n\neq i)}}^{N} w_{in} e^{j\varphi_{p(i)}^{tx}}$ |
| Tx phase and rx phase must be conjugate $\varphi_{p(i)}^{tx} = \varphi_{p(i)}^{rx}{}^*$ | Tx phase $\varphi_{p(i)}^{tx}$ and rx phase $\varphi_{p(i)}^{rx}$ not correlated |
| Cluster set $C = \{e^{j\varphi_k}\}_{k=1}^{K}$ | Phase set $P = \left\{e^{j\varphi_k^{tx}}\right\}_{k=1}^{K}$ |

## C. ASSUMPTIONS

Considering the proposed surrogate-symbol based co-channel wireless system in FIGURE **4**, we assume the following:

- Assumption-1 (A1): Each receiver correctly estimates the baseband channel coefficient, Channel State Information (CSI), from each co-channel transmitter.
- Assumption-2 (A2): Each receiver reports these CSIs to a centralized unit (which can possibly be one of the transmitters).

- Assumption-3 (A3): The “genie” centralized unit can process the instantaneous channel gain matrix **H** and can send the transmit phase index to corresponding transmitter.
- Assumption-4 (A4): Each tx sends the rx-phase index ($i.e.$, $\varphi_{p(i)}^{rx}$) to the corresponding rx from a separate and CCI-free channel.
- Assumption-5 (A5): The previous steps 1 to 4 are processed within the channel coherence time.

Throughout this paper, the channel matrix **H** is not necessarily the raw physical channel matrix. Power control, CoMP, and any other radio resource management or interference-reduction algorithm operate at a different layer and shape the channel matrix **H** itself. We therefore assume that **H** in this paper refers to the *effective* channel matrix obtained after any such algorithm has already been applied, not the absolute channel matrix. Consequently, the gain promised by the proposed algorithms comes on top of any other CCI reduction technique, as our framework here is based on a bottom-line baseband model that takes the already-optimized channel as its starting point.

The proposed method in this paper under the assumptions above are depicted in FIGURE 5.

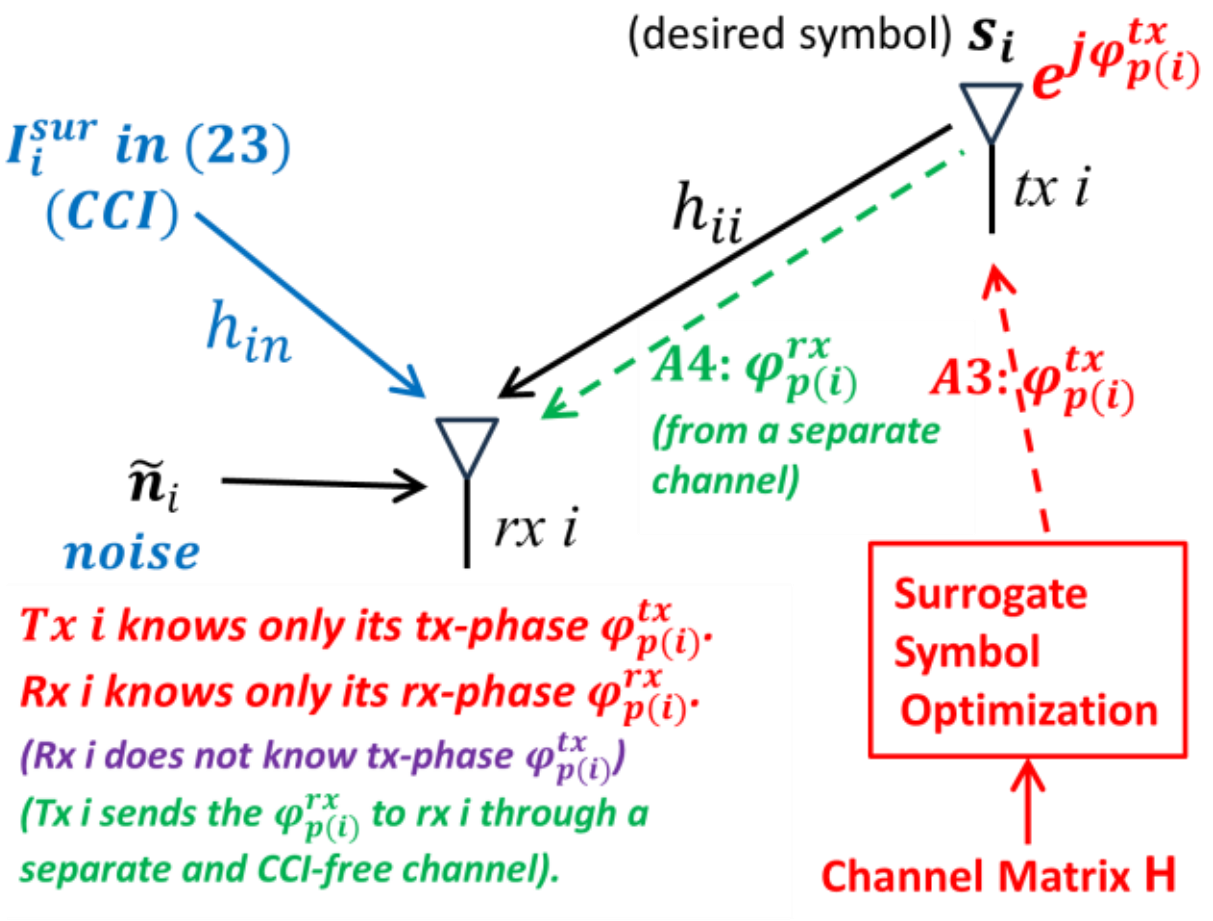

**FIGURE 5. The proposed method in the eyes of the rx *i***

*In the proposed method in FIGURE 5, it is worth to emphasize that the surrogate symbol optimization is done using only and only the channel matrix **H** and discrete transmit phase set $\{\varphi_p^{tx}\}_{p=1}^{p=K}$, and nothing else (i.e., without using the receiver phase set $\{\varphi_p^{rx}\}_{p=1}^{p=K}$, and without the desired signal $s_i$). Futhermore, the receiver i has no information about the transmit phase i ($\varphi_{p(i)}^{tx}$, surrogate symbol i) at all.*

The CSI reporting in cellular radio systems has been widely studied for over three decades by both academia and industry. A large body of literature, including numerous research articles and patents, has addressed the topics underlying most of the assumptions adopted in this work. The practical realization of these assumptions in operational real cellular systems is beyond the scope of this paper. A selection of representative patents addressing our assumptions is provided in the Appendix for completeness.

In this paper, under the above assumptions A1 to A5, we present a novel optimization problem, called “Game of Phases – Surrogate Symbol Optimization”, and present two algorithms to solve it. In this paper, we focus on single-antenna case.

**Perfect CSI as an upper bound.** Assumption A1 follows the standard practice of establishing performance limits of multiuser and interference networks under perfect CSI [32]-[33]. In a practical system, both the surrogate phases chosen by the centralized unit and the de-rotation at each receiver would be computed from an estimate of **H** rather than from **H** itself. The SER and BER results in Section VI are therefore an upper bound on the performance (i.e., a lower bound on the error rates) that a practical system with channel-estimation error can achieve. The comparison is nevertheless fair: The traditional system in Fig.2, against which all gains are measured, is also evaluated under perfect CSI, since its receiver uses the exact own-channel phase. Both systems are therefore compared at their best case, under identical channel knowledge. Whether the gain of the proposed algorithms is preserved under channel-estimation error is left to future work.

### D. GAME OF PHASES: DETERMINING OPTIMUM SURROGATE SYMBOLS IN WIRELESS COMMUNICATIONS

In this section, we propose a novel optimization problem, called “Game of Phases” as follows: How can we optimally determine the Surrogate Symbols which minimize or maximize a cost function representing the overall $N$ co-channel wireless system performance metric (e.g. average interference power, or average SINR, or average BER, etc) in FIGURE 4? To be precise, the novel optimization problem is defined as follows: Given an $N$-by-$N$ channel matrix **H**, and given a discrete set of phases $\{\varphi_p^{tx}\}_{p=1}^{p=K}$, what are the optimum Surrogate Symbols which minimize/maximize a pre-defined wireless performance metric denoted as $J_{per}(\cdot)$, i.e.,

$$\arg \max_{\{\varphi_{p(n)}^{tx}\}_{i=1}^{N}} J_{per}\left(\mathbf{H}, \{\varphi_{p(n)}^{tx}\}_{i=1}^{N}\right) \qquad (27)$$

The proposed optimization problem is depicted in FIGURE 6.

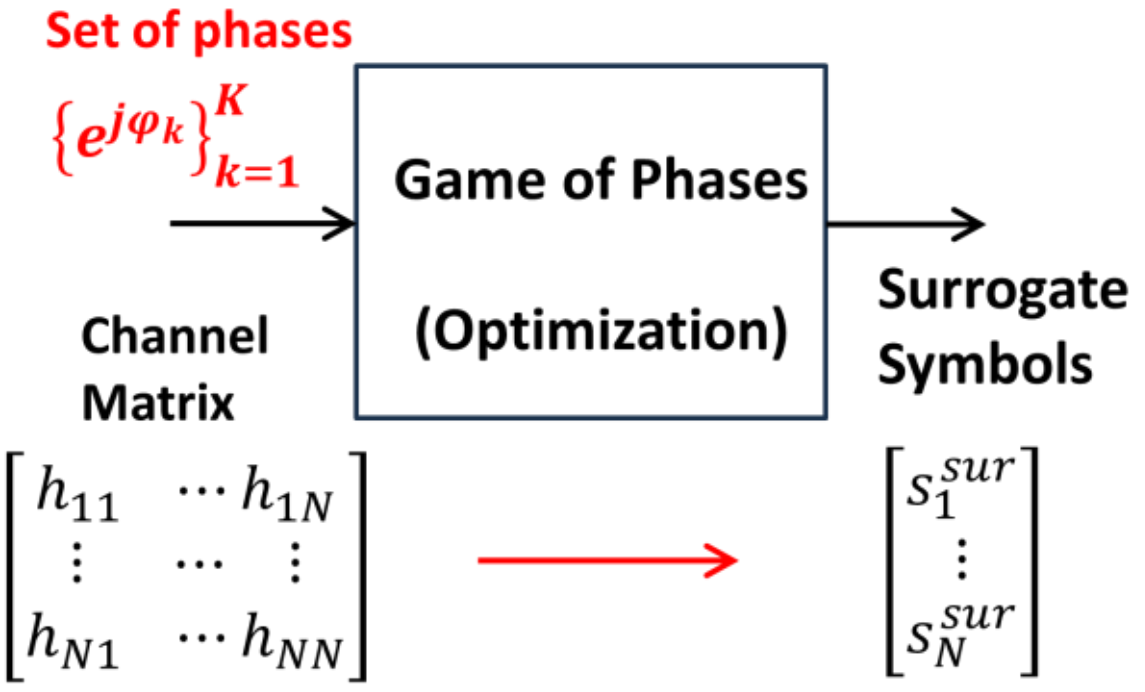

**FIGURE 6. Proposed optimization problem: Determining Optimum Surrogate Symbols in Wireless Communications.**

## IV. PROPOSED ALGORITHMS: MIPIP AND MAPRIS

In this section, in order to solve the "optimum surrogate symbols problem" defined in **FIGURE 6**, we present two novel phase optimization algorithms, which we call "Minimizing Phased-Interference Power" (MiPIP) algorithm and "Maximizing Phased-and-Rotated Interference Support" (MaPRIS) algorithm.

### A. MIPIP (MINIMUM PHASED-INTERFERENCE POWER ALGORITHM

Being inspired from [2], we introduce the *phased*-CCI as follows:

**Definition**: ***Phased* CCI**: We define the *phased* CCI for receiver *i*, denoted as $\boldsymbol{I}_i^{ph}$, as follows

$$\boldsymbol{I}_i^{ph}(\{\boldsymbol{h}_{in}, \boldsymbol{s}_n^{sur}\}_{n=1}^N) = e^{-j\theta_{ii}} \boldsymbol{I}_i^{sur} \tag{28}$$

$$= e^{-j\theta_{ii}} \sum_{\substack{n=1 \\ (n \neq i)}}^{N} \boldsymbol{s}_n^{sur} \boldsymbol{h}_{in}$$

$$= \sum_{\substack{n=1 \\ (n \neq i)}}^{N} c_n\, \alpha_{in} e^{j\left(\varphi_{p(n)}^{tx} + \theta_{in} - \theta_{ii}\right)}$$

where $\boldsymbol{s}_n^{sur} = c_n e^{j\varphi_{p(n)}^{tx}}$ is the surrogate symbol of tx $n$, $\boldsymbol{I}_i^{sur}$ is defined in (23) and $\{\varphi_p^{tx}\}_{p=1}^{p=K}$. From (24) and (28), we note that $\boldsymbol{I}_i^{ph}$ is the CCI part of the signal entering the MLD in **FIGURE 4** for rx $i$. From (28), $|\boldsymbol{I}_i^{sur}| = |\boldsymbol{I}_i^{ph}|$. So, we define the cost function of the MiPIP algorithm denoted as $J_{MiPIP}$ as the sum (or average) of the absolute values of the phased CCI:

$$J_{MiPIP}(\mathbf{H}, \{\boldsymbol{s}_n^{sur}\}_{n=1}^N) = \sum_{i=1}^{N} \left| \boldsymbol{I}_i^{sur}\left(\left\{c_n \varphi_{p(n)}^{tx}\right\}_{n=1}^N\right) \right| \tag{29}$$

$$= \sum_{i=1}^{N} \left| \sum_{\substack{m=1 \\ (m \neq i)}}^{N} \alpha_{im} \boldsymbol{s}_m^{sur} e^{j\theta_{im}} \right|$$

where $\boldsymbol{s}_n^{sur} = c_n \varphi_{p(n)}^{tx}$ and $\{\varphi_p^{tx}\}_{p=1}^{p=K}$. So, the proposed MiPIP algorithm suggests to solve the "game of phases" optimization problem depicted in FIGURE 6 by minimizing the $J_{MiPIP}(\mathbf{H}, \{\boldsymbol{s}_n^{sur}\}_{i=1}^N)$ in (29), i.e.,

$$min_{\{s_i^{sur}\}_{i=1}^N} \; J_{MiPIP}(\mathbf{H}, \{\boldsymbol{s}_i^{sur}\}_{i=1}^N) \tag{30}$$

where $\boldsymbol{s}_i^{sur} = c_i \varphi_{p(i)}^{tx}$. In other words, the optimum surrogate symbols $\{\boldsymbol{s}_i^{sur}\}_{i=1}^N$ are determined by the optimization in (30). The asynchronous, one-node-at-a-time update strategy adopted by MiPIP and MaPRIS follows the same relaxation principle used in Monte Carlo methods for finding low-energy configurations of discrete-state systems [10], where sequentially updating individual sites according to a local decision rule drives the overall system toward a stationary point of the global cost function. The proposed MiPIP algorithm is summarized in Table 3. For the sake of brevity, in the rest of the paper, we assume $c_i = 1$. Even under the assumption $c_i = 1$, the phase optimization problem $\{\varphi_{p(i)}^{tx}\}_{i=1}^N$ is already a challenging one.

**Table 3 The MiPIP algorithm**

| |
|---|
| The surrogate symbols are sequentially (asynchronously) updated as follows: |
| Step (1) : Determine the winning surrogate symbol for tx $i$ at time $t$ in such a way that tx $i$ reduces the average phased CCIs. $$w(i) = index_{\,p(i)}\; min\left\{ \sum_{n=1}^{N} \left|\boldsymbol{I}_{\boldsymbol{n}}^{sur}\left(\varphi_{p(i)}^{tx}\right)\right| \right\} \tag{31}$$ where $w(i)$ represents the index of the winning surrogate symbol (i.e, the winning tx phase) among the set $\{\varphi_{p(i)}^{tx}\}_{p(i)=1}^{p(i)=K}$ for tx $i$. |
| Step (2) : Continue step (1) *asynchronously* for each tx one by one until there is no surrogate symbol change any more, reaching a local minima of the MiPIP cost function $J_{MiPIP}$ in (29). |

**Proposition 2:** Consider the proposed surrogate-symbol based co-channel wireless system in **FIGURE 4**. Let $\boldsymbol{H} \in C^{N \times N}$ be an arbitrary channel matrix (complex-valued matrix). The proposed MiPIP algorithm in (31) above <u>*minimizes*</u> the sum (or average) of the absolute values of the phased CCI defined in (29).

***Proof:*** The proof is straightforward. Whenever a tx is assigned from its current surrogate symbol to a different surrogate symbol by (31), the $J_{MiPIP}(\mathbf{H}, \{\boldsymbol{s}_n^{sur}\}_{i=1}^N)$ is strictly decreased due to the *min* operation. In other words, whenever a surrogate symbol is changed, $J_{MiPIP}(t+1) < J_{MiPIP}(t)$, where $t$ is step number. Because the minimum of $J_{MiPIP}(t)$ in (29) is finite, the MIPIP in Table 3 converges to a local minima after a finite number of steps, which completes the proof. ■

***Corollary 2:*** Let $\boldsymbol{H} \in C^{N \times N}$ be an arbitrary channel matrix (complex matrix). Comparing the traditional wireless system in **FIGURE 2** and the proposed surrogate-symbol based wireless system in FIGURE **4** using the MiPIP algorithm, we conclude that the MiPIP algorithm reduces the BER simply because the power (i.e. variance) of the average interference is reduced due to the min operation in (31).

### B. MAPRIS (MAXIMIZING PHASED-AND-ROTATED INTERFERENCE SUPPORT)

Before presenting our second algorithm, we first define the *phased and rotated CCI* as follows:

**Definition**: ***Phased-and-Rotated* CCI ($\boldsymbol{I}_i^{ph-r}$)**: From (25), we define the interference entering the MLD of the rx $i$ in our proposed wireless system in **FIGURE 4**, as the *phased-and-rotated* CCI for receiver *i*. It is denoted as $\boldsymbol{I}_i^{ph-r}$ and is given as follows (from (25)):

$$\boldsymbol{I}_i^{ph-r} = e^{-j\varphi_{p(i)}^{rx}} \boldsymbol{I}_i^{ph} \quad (32)$$
$$\boldsymbol{I}_i^{ph-r} = e^{-j\varphi_{p(i)}^{rx}} e^{-j\theta_{ii}} \boldsymbol{I}_i^{sur}$$

where $\boldsymbol{I}_i^{ph}$ is the *phased* CCI of rx *i* as defined in (28) and $\boldsymbol{I}_i^{sur}$ is defined in (23). Let's represent the CCI signal $\boldsymbol{I}_i^{sur}$ in (32) as follows:

$$\boldsymbol{I}_i^{sur} = e^{j\vartheta_i} |\boldsymbol{I}_i^{sur}| \quad (33)$$

where $\vartheta_i$ is the angle of $\boldsymbol{I}_i^{sur}$ for rx *i*.

In what follows, we first present our second algorithm to solve the "optimum surrogate symbols problem" (defined in **FIGURE 6**), and then we present its analysis: Our algorithm is called "*Maximizing Phased-and-Rotated Interference Support*" (MaPRIS). We define its cost function, $J_{MaPRIS}$, to be maximized as follows:

$$J_{MaPRIS}(\mathbf{H}, \{\boldsymbol{s}_i^{sur}\}_{i=1}^{N}) = \sum_{i=1}^{N} \cos(\vartheta_i - \theta_{ii} - \varphi_{p(i)}^{tx}) \quad (34)$$

or alternatively

$$\hat{J}_{MaPRIS}(\mathbf{H}, \{\boldsymbol{s}_i^{sur}\}_{i=1}^{N}) = \sum_{n=1}^{N} |\vartheta_i - \theta_{ii} - \varphi_{p(i)}^{tx}| \quad (35)$$

to be minimized, where $\vartheta_i$, $\theta_{ii}$ and $\varphi_{p(i)}^{tx}$ is the phase (angle) of the surrogate interference $\boldsymbol{I}_i^{sur}$, phase of the channel coefficient $\boldsymbol{h}_{ii}$ and the surrogate symbol phase of tx *n*, respectively, and $\boldsymbol{s}_i^{sur} = c_i\varphi_{p(i)}^{tx}$. The proposed MaPRIS algorithm suggests to solve the "game of phases" optimization problem by maximizing $J_{MaPIS}$ in (34), i.e.,

$$\max_{\{\varphi_{p(i)}^{tx}\}_{i=1}^{N}} J_{MaPIS}\left(\mathbf{H}, \{\varphi_{p(i)}^{tx}\}_{i=1}^{N}\right) \quad (36)$$

or alternatively, minimizing $\hat{J}_{MaPIS}$ in (35), i.e.,

$$\min_{\{\varphi_{p(i)}^{tx}\}_{i=1}^{N}} \hat{J}_{MaPRIS}\left(\mathbf{H}, \{\varphi_{p(i)}^{tx}\}_{i=1}^{N}\right) \quad (37)$$

where $\{\varphi_p^{tx}\}_{p=1}^{p=K}$. The proposed MaPRIS algorithm is summarized in **Table 4**. In what follows, we use the cost function $J_{MaPRIS}$ in (34) for the sake of brevity.

**Table 4 The MaPRIS algorithm ([3])**

| |
|---|
| The surrogate symbols are sequentially (asynchronously) updated as follows: |
| Step (1) : Determine the set of candidate surrogate symbols for tx *i*, represented as $S_w^i$ as follows: $S_w^i = indices_s\{\cos(\vartheta_i - \theta_{ii} - \varphi_s) < \cos(\vartheta_i - \theta_i - \varphi_{p(i)}^{tx})\}$ (38) |
| Step (2) : If $S_w^n$ is not an empty set, then the surrogate symbol of tx *n* is updated at time *(t+1)* according to the following simple rule $w_i = index_{p\in S_w^n} \max\left\{\sum_{n=1}^{N} \cos(\vartheta_i - \theta_{ii} - \varphi_{p(i)}^{tx})\right\}$ (39) where $w_i$ represents the index of the winning surrogate symbol (i.e, the tx phase) among the set $S_w^i$ . |
| Step (3) : Continue step (1) and (2) *asynchronously* for each tx one by one until there is no surrogate symbol change any more, reaching a local maxima of the MaPRIS cost function $J_{MaPIS}$ in (34) or $\hat{J}_{MaPIS}$ in (35). |

Before presenting the analysis of the MaPRIS, we first define the following:

**Definition**: ***Effective Angle*** **($\boldsymbol{\Xi}_i^{eff}$)**: We define the argument of the cosine in (38) and (39) in the MaPRIS algorithm as the ***Effective Angle*** for rx *i*, i.e.,

$$\boldsymbol{\Xi}_i^{eff} = \vartheta_i - \theta_{ii} - \varphi_{p(i)}^{tx} \quad (40)$$

**Definition**: ***Rotating Angle*** **($\boldsymbol{\gamma}_i^{rot}$)**: From (32), we define the angle by which $\boldsymbol{I}_i^{sur}$ (the *phased* CCI) is rotated as the ***Rotating Angle*** for rx *i*, and is denoted as $\boldsymbol{\gamma}_i^{rot}$, i.e.,

$$\boldsymbol{\gamma}_i^{rot} = \theta_{ii} + \varphi_{p(i)}^{tx} \quad (41)$$

From (40) and (41),

$$\boldsymbol{\Xi}_i^{eff} = \vartheta_i - \gamma_i^{rot} \quad (42)$$

In order to give an insight into how the MaPRIS algorithm evolves by steps, we sketch a random snapshot before and after the tx-phase update of (39) in FIGURE 7.*(a)* and *(b)*, respectively. In FIGURE 7, $\boldsymbol{\Xi}_i^{eff(1)}$ and $\boldsymbol{\Xi}_i^{eff(2)}$ represents the *effective angle* before and after the tx-phase update of (39), respectively. Similarly, $\boldsymbol{s}_i^{sur(1)}$ and $\boldsymbol{s}_i^{sur(2)}$ represents the surrogate symbol before and after the tx-phase update, respectively, and $\boldsymbol{\gamma}_i^{rot(1)}$ and $\boldsymbol{\gamma}_i^{rot(2)}$ is the rotating angle before and after the tx-phase update, respectively.

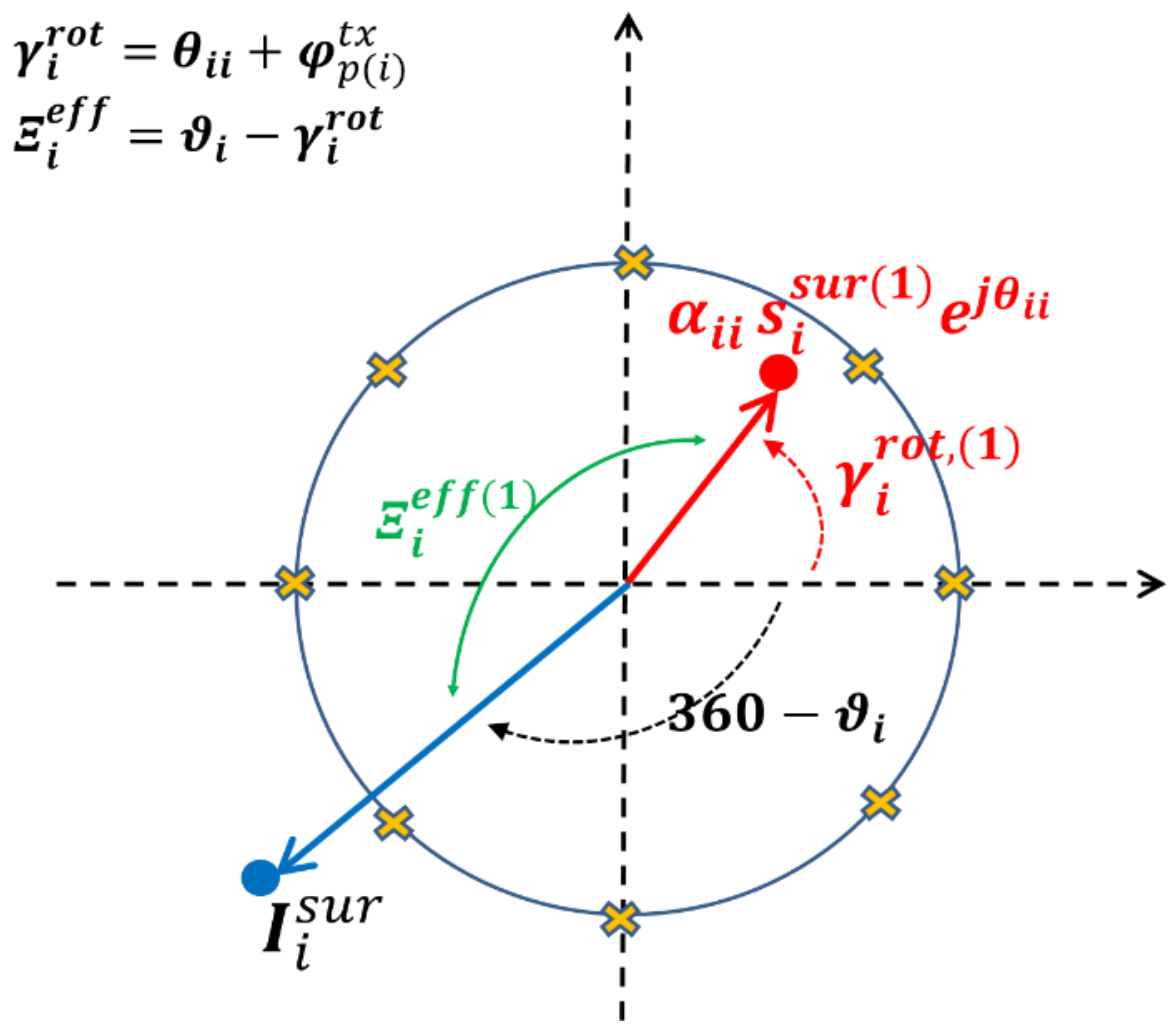

*(a)* before *tx*-phase update

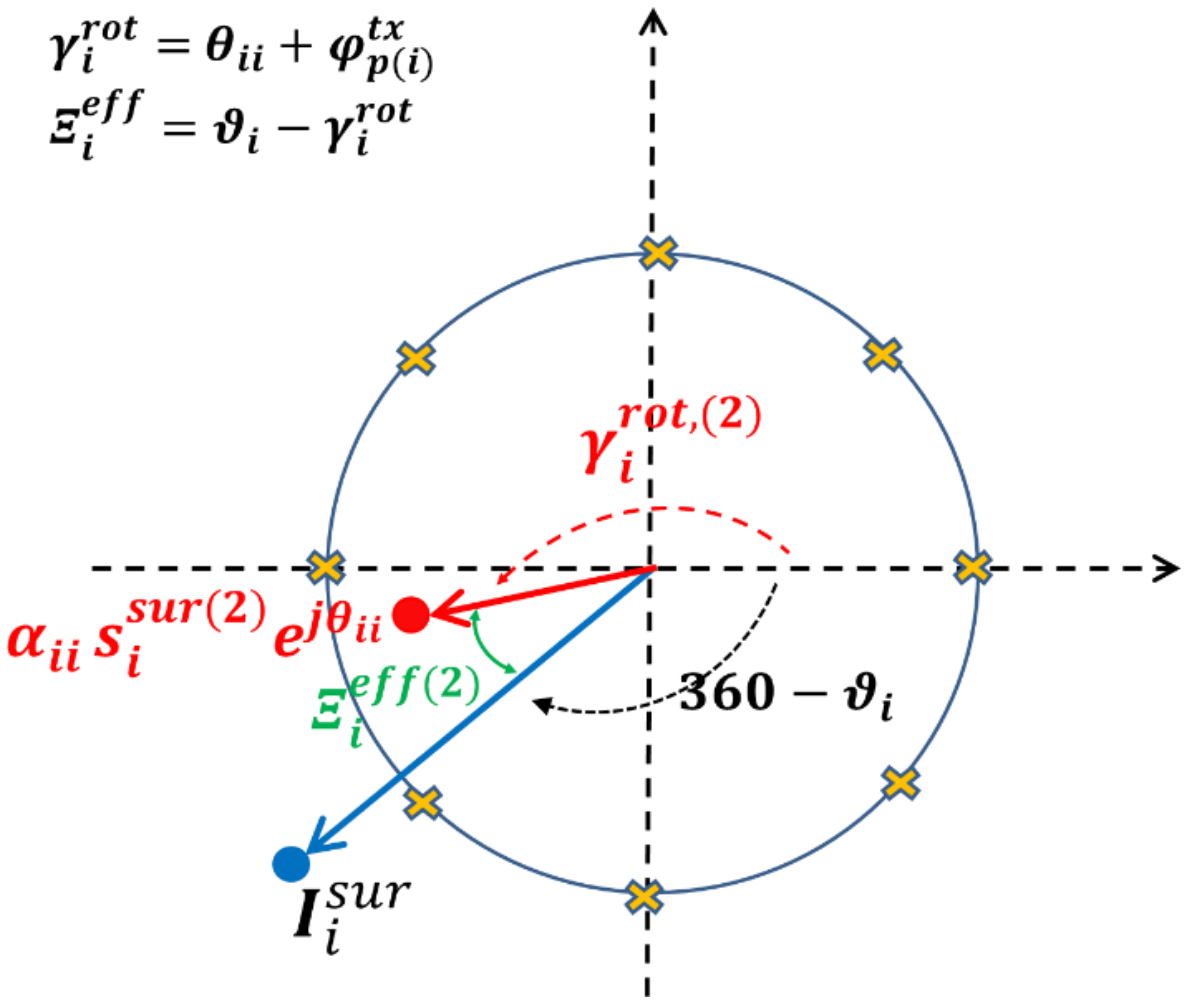

(*b*) after *tx*-phase update

**FIGURE 7. Illustration of the *effective angle* $\Xi_i^{eff}$, surrogate symbol $s_i^{sur}$ , and rotating angle $\gamma_i^{rot}$ in the MaPRIS in (39). *(a)* before phase update, and *(b)* after phase update.**

**Proposition 3:** Consider the proposed surrogate-symbol based co-channel wireless system in FIGURE 4. Let $H \in C^{N\times N}$ be an arbitrary channel matrix (complex-valued matrix). The proposed MaPRIS algorithm in Table 4 above *maximizes* the average cosine of the *effective angle*, i.e.,

$$J_{MaPRIS}\left(\mathbf{H}, \{\varphi_{p(i)}^{tx}\}_{i=1}^{N}\right) = \frac{1}{N}\sum_{i=1}^{N} cos\left(\Xi_i^{eff}\right) \tag{43}$$

which is equal to $J_{MaPIS}$ in (34).

***Proof:*** Whenever a tx is assigned from its current surrogate symbol to a different surrogate symbol by (39) in Table 4, the $J_{MaPIS}\left(\mathbf{H}, \{\varphi_{p(i)}\}_{i=1}^{N}\right)$ is strictly increased due to the *max* operation in (39). In other words, whenever a surrogate symbol is changed, $J_{MaPIS}(t+1) > J_{MaPIS}(t)$. Because the maximum of $J_{MaPRIS}(t)$ in (34) or (43) is finite, the MaPRIS converges to a local maxima after a finite number of steps, which completes the proof. ■

The CCI part of the input signal to the MLD of the rx *i* in FIGURE 4 is the *phased-and-rotated* CCI $I_i^{ph-r}$ in (32).

**Proposition 4:** Consider the proposed surrogate-symbol based
wireless system in FIGURE **4** using the MaPRIS algorithm in Table 4. We conclude the following: The MaPRIS converts the "hostile" CCI into a "friendly signal" which is "almost" the desired symbol. In other words, the MaPRIS algorithm makes the *phased-and-rotated* CCI $I_i^{ph-r}$ in (32) look like the desired symbol $s_i$ to be transmitted as much as possible using the *K* discrete phase options at the transmitter. This is because the *Effective Angle* $\Xi_i^{eff}$ turns out to be also the angle between $I_i^{ph-r}$ and the desired symbol $s_i$.

***Proof:*** Using (14), (32) and (33) in (25), we obtain the signal entering the MLD in FIGURE **4** as

$$\tilde{\tilde{s}}_i = \alpha_{ii} \underbrace{\underbrace{c_i e^{j\varphi_{p(i)}^{tx}}}_{=s_i^{sur}} e^{-j\varphi_{p(i)}^{rx}}}_{=s_i} + I_i^{ph-r} + e^{-j\left(\varphi_{p(i)}^{rx}+\theta_{ii}\right)} n_i \tag{44}$$

$$\tilde{\tilde{s}}_i = \alpha_{ii} s_i + I_i^{ph-r} + \tilde{n}_i$$

where $s_i$ is the desired symbol to be transmitted and $I_i^{ph-r} = e^{j\left(\vartheta_i - \theta_{ii} - \varphi_{p(i)}^{rx}\right)}|I_i^{sur}|$, and $\tilde{n}_i = e^{-j\left(\varphi_{p(i)}^{rx}+\theta_{ii}\right)} n_i$ thermal noise.

From (44), the angle between the desired symbol $s_i$ and the *phased-and-rotated* interference vector $I_i^{ph-r}$ is

$$\underbrace{\left(\vartheta_i - \theta_{ii} - \varphi_{p(i)}^{rx}\right)}_{\text{angle of } I_i^{ph-r}} - \underbrace{\left(\varphi_{p(i)}^{tx} - \varphi_{p(i)}^{rx}\right)}_{\text{angle of desired } s_i} \tag{45}$$

$$= \vartheta_i - \theta_{ii} - \varphi_{p(i)}^{tx}$$
$$= \Xi_i^{eff}$$

the *effective Angle* $\Xi_i^{eff}$ as defined in (40).

Eq.(45) implies that the smaller the $\Xi_i^{eff}$ is, the more $I_i^{ph-r}$ looks like the desired signal $s_i$. Furthermore, because the cost function of the MaPRIS algorithm in (34) or (35) minimizes the effective angle $\Xi_i^{eff}$, we conclude that the MaPRIS algorithm makes the *phased-and-rotated* CCI $I_i^{ph-r}$ in (32) look like the desired symbol $s_i$ to be transmitted as much as possible using the *K* discrete phase options at the transmitter and receiver. In other words, the MaPRIS converts the "hostile" CCI into a "friendly signal" which is close-to the desired symbol as much as possible limited by the *K* discrete phase options. This completes the proof. ■

In order to give an insight into Proposition 4 and eq.(45), we sketch $I_i^{ph-r}$, $s_i$ and $\Xi_i^{eff}$ for a random channel realization in **FIGURE 8**.

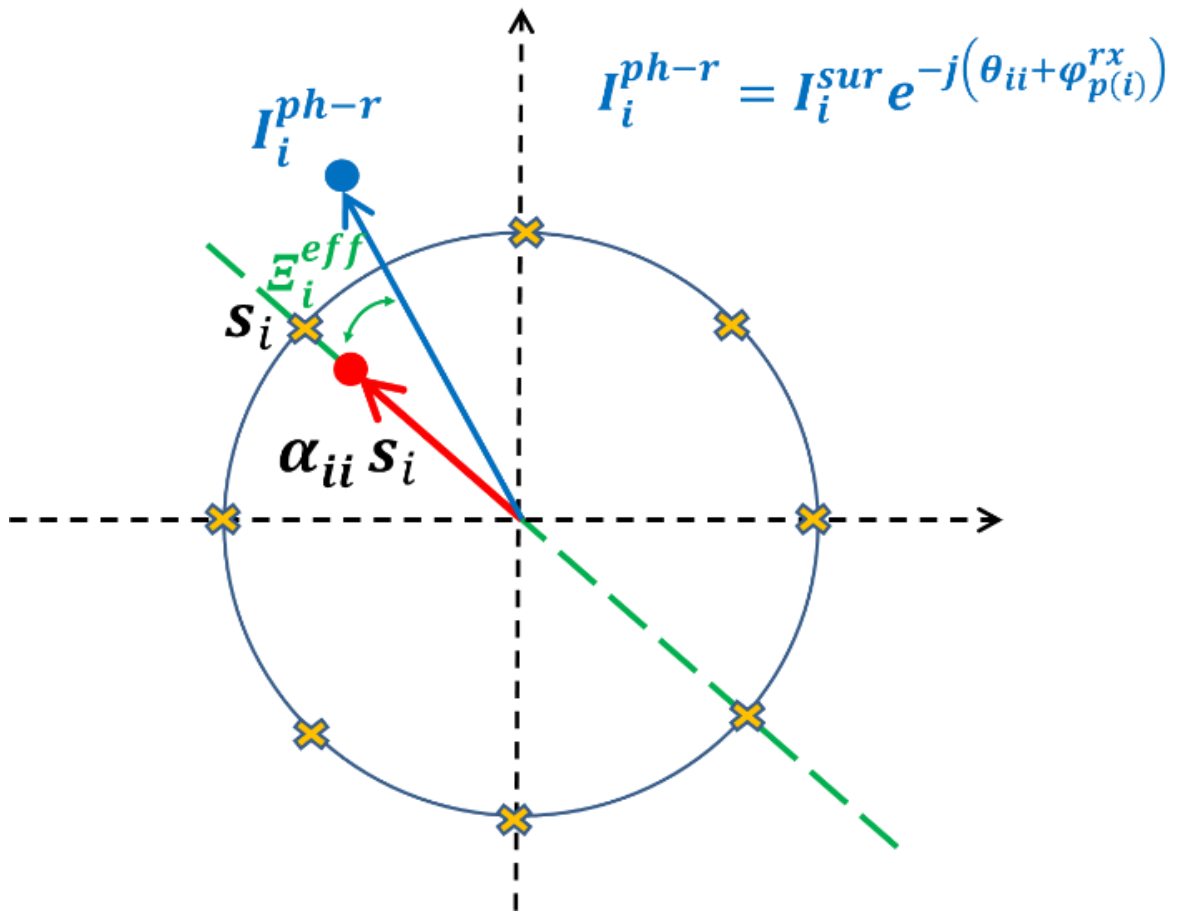

**FIGURE 8. Illustration of phased and rotated CCI $I_i^{ph-r}$, desired symbol $s_i$ and the *effective angle* $\Xi_i^{eff}$ at receiver *i* for a random channel realization.**

**Proposition 5:** Let $\boldsymbol{H} \in C^{N \times N}$ be an arbitrary channel matrix (complex-valued matrix). As compared to the traditional wireless system in **FIGURE 2**, the proposed surrogate-symbol based wireless system in FIGURE **4** using the MaPRIS algorithm in Table 4 (drastically) reduces the BER. This follows from the following counter-intuitive finding: In FIGURE **4** with the MaPRIS, the higher the CCI variance is, the higher the SINR is, and equivalently, the higher the data rate is.

***Proof:*** Before we delve into the proof, we first define the following:

**Definition**: **"Interference Support":** We define "interference support", denoted as $\boldsymbol{I}_i^{sup}$, as the projection of the *phased and rotated interference* vector $\boldsymbol{I}_i^{ph-r}$ in (32) onto the desired symbol vector $\boldsymbol{s}_i$:

$$\boldsymbol{I}_i^{sup} = |\boldsymbol{I}_i^{ph-r}|\cos(\boldsymbol{\Xi}_i^{eff})\boldsymbol{s}_i \tag{46}$$

The "Interference Support" vector is depicted in FIGURE 9.

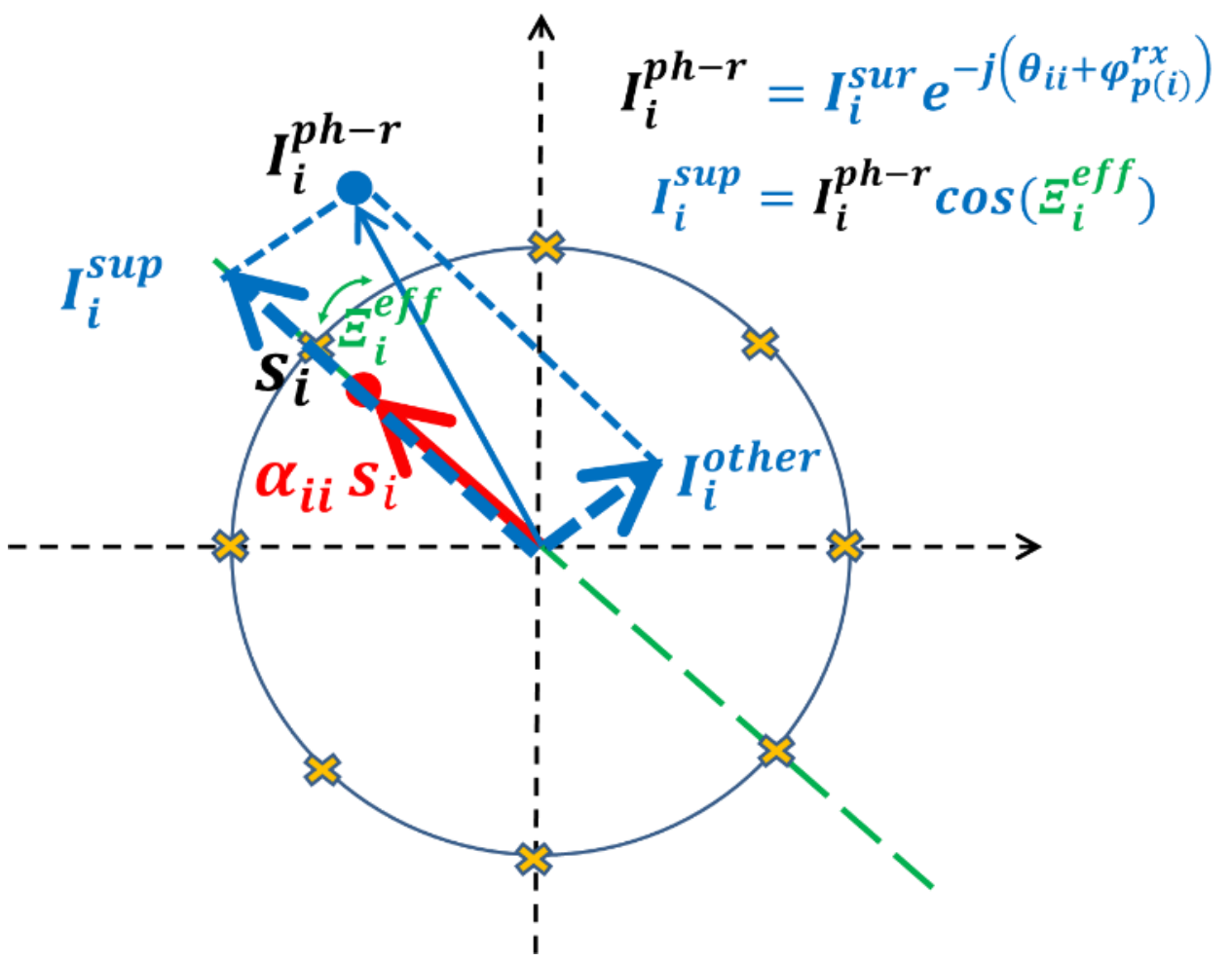

**FIGURE 9. Illustration of the "interference support" $I_i^{sup}$.**

So, from FIGURE 9, we write

$$\boldsymbol{I}_i^{sup} = \boldsymbol{I}_i^{sup} + \boldsymbol{I}_i^{other} \tag{47}$$

where $\boldsymbol{I}_i^{other}$ is perpendicular vector to the $\boldsymbol{I}_i^{sup}$.

From FIGURE 9, because both the $\boldsymbol{I}_i^{sup}$ and $\boldsymbol{s}_i$ are two vectors in the same direction, let's denote the "Interference Support" vector $\boldsymbol{I}_i^{sup}$ as

$$\boldsymbol{I}_i^{sup} = \beta_i\, \boldsymbol{s}_i \tag{48}$$

where $\beta_i = |\boldsymbol{I}_i^{ph-r}|\cos(\boldsymbol{\Xi}_i^{eff}) > 0$ is the magnitude of the Interference Support vector.

Using (32) and (33) in (25), and (48)-(47), we obtain the signal entering the MLD in FIGURE 4 as

$$\tilde{\tilde{\boldsymbol{s}}}_i = (\alpha_{ii} + \beta_i)\boldsymbol{s}_i + \boldsymbol{I}_i^{other} + \tilde{\boldsymbol{n}}_i \tag{49}$$

On the other hand, the signal entering the MLD in the traditional co-channel wireless system in **FIGURE 2** is given by (11):

$$\tilde{\boldsymbol{s}}_i = \alpha_{ii}\boldsymbol{s}_i + e^{-j\theta_{ii}}\boldsymbol{I}_i + e^{-j\theta_{ii}}\boldsymbol{n}_i \tag{50}$$

Comparing (49) with (50), one obtains

$$\begin{gathered} \alpha_{ii} + \beta_i > \alpha_{ii} \\ |\boldsymbol{I}_i^{other}| < |\boldsymbol{I}_i^{ph-r}|, \text{ and} \\ |\tilde{\boldsymbol{n}}_i| = |e^{-j\theta_{ii}}\boldsymbol{n}_i| = |\boldsymbol{n}_i|, \end{gathered} \tag{51}$$

From (51), the desired symbol power is increased and the CCI power is decreased while the thermal noise power remains the same by the MaPRIS algorithm. In other words, the higher the CCI variance is, the higher the SINR is. This implies that the MaPRIS reduces the BER as compared to the traditional wireless system in **FIGURE 2**. This completes the proof. ■

**Corollary 3:** Because phase or rotation does not change the norm of a vector, i.e., the *phased-and-rotated* CCI $\boldsymbol{I}_i^{ph-r}$ in (32) has the same norm as the surrogate interference $\boldsymbol{I}_i^{sur}$, i. e. $|\boldsymbol{I}_i^{ph-r}|=|\boldsymbol{I}_i^{sur}|$. The proposed MaPRIS algorithm in Table 4 using (34) maximizes the sum (or average) of the Interference Supports $\boldsymbol{I}_i^{sup}$ in (46), i.e.,

$$\text{MaPRIS with (34):} \quad max_{\{\varphi_{p(i)}^{tx}\}_{i=1}^{N}} \frac{1}{N}\sum_{i=1}^{N}|\boldsymbol{I}_i^{sup}| \tag{52}$$

If the cost function $\hat{J}_{MaPRIS}$ in (35) is used, then the MaPRIS minimizes the sum (or average) of $|\boldsymbol{\Xi}_i^{eff}|$

$$\text{MaPRIS with (35):} \quad min_{\{\varphi_{p(i)}^{tx}\}_{i=1}^{N}} \frac{1}{N}\sum_{i=1}^{N}|\boldsymbol{\Xi}_i^{eff}| \tag{53}$$

### *C. SUMMARY*

*As a quick overall summary: The MaPRIS does not directly reduce the phased interference power (i.e., after all the interference variance remains the same), but instead it aims to maximize the "support" of the phased and rotated interference* $\boldsymbol{I}_i^{ph-r}$ *on the desired symbol (power). The MaPRIS achieves this by minimizing the angle between the rotated surrogate symbol vector and the surrogate interference vector <u>at the centralized unit </u>(which can be one of the transmitters). This very angle turns out to be equal to the angle between the desired symbol and the phased and rotated interference vector <u>at the receiver</u>. This is the key merit of the MaPRIS which converts the hostile co-channel interference into a "friendly signal" which is close-to the desired symbol. In other words this key merit makes the phased and rotated interference* $\boldsymbol{I}_i^{ph-r}$ *"looks like" the desired symbol* $\boldsymbol{s}_i$ *as much as possible just before the MLD.*

*<u>It is worth to note that the MaPRIS achieves these results presented above using only and only the channel matrix **H** and discrete transmit phase set</u>* $\{\varphi_p^{tx}\}_{p=1}^{p=K}$*<u>, and nothing else (i.e., without using the rx phase set</u>* $\{\varphi_p^{rx}\}_{p=1}^{p=K}$*<u>, and without the</u>*

*desired signal $s_i$). Futhermore, the receiver i has no information about the transmit phase i ($\varphi^{tx}_{p(i)}$) at all.*

Last, but not least, while the cost function $J_{MaPRIS}$ in (34) maximizes the average $cos(\Xi_i^{eff})$, or equivalently the average of $|I_i^{sup}|$ as shown in (52), the $\hat{J}_{MaPRIS}$ in (35) minimizes the average of the angle $|\Xi_i^{eff}|$ itself as shown in (53).

### D. COMPUTATIONAL COMPLEXITY

We analyze the computational complexity of MiPIP and MaPRIS for an $N$ co-channel tx-rx pairs network with a discrete phase set of size $K$. Both algorithms follow the same asynchronous update structure (Table 3, Table 4): at each step, one transmitter evaluates its cost function across all K candidate phases, and this is repeated, tx by tx, until convergence.

Table 5 summarizes the computational complexity. Because MiPIP and MaPRIS share the same sum-over-$N$, compare-over-$K$, repeat-over-$N$ update structure, they are asymptotically identical in both time and memory complexity.

**Table 5 Computational complexity of MiPIP and MaPRIS.**

| Quantity | MiPIP and MaPRIS |
|---|---|
| One candidate-phase evaluation | $O(N)$ |
| One transmitter's decision ($K$ candidates) | $O(N \cdot K)$ |
| One full sweep ($N$ transmitters) | $O(N^2 \cdot K)$ |
| Total to convergence ($T$ sweeps) | $O(T \cdot N^2 \cdot K)$ |
| Memory | $O(N^2)$ for **H**, $O(N)$ for phase state |

## V. RESOLVING THE THOUGHT EXPERIMENT: RECOVERING THE DESIRED SYMBOL UNDER A FULLY-BLOCKED OWN CHANNEL

In this section we resolve, in full mathematical detail, the "thought experiment" introduced in Section I. . We show that the general framework developed in Sections III. –IV. (which is derived there for the case of a nonzero own-channel link) extends to the most extreme and seemingly hopeless scenario as a special case: one in which *every* receiver's own desired channel link is simultaneously zero.

### A. THE THOUGHT EXPERIMENT, REVISITED

Consider a co-channel interference wireless scenario in which the own-signal channel is fully blocked for every receiver, i.e.,

$$\boldsymbol{h}_{ii} = \alpha_{ii} e^{j\theta_{ii}} = 0, \qquad i = 1, \ldots, N \tag{54}$$

while all interfering channels $h_{in}$, $n \neq i$, remain unchanged.

This converts the general system of FIGURE 4 into the system shown in FIGURE 10, where $h_{ii} = 0$ (there is no channel to estimate).

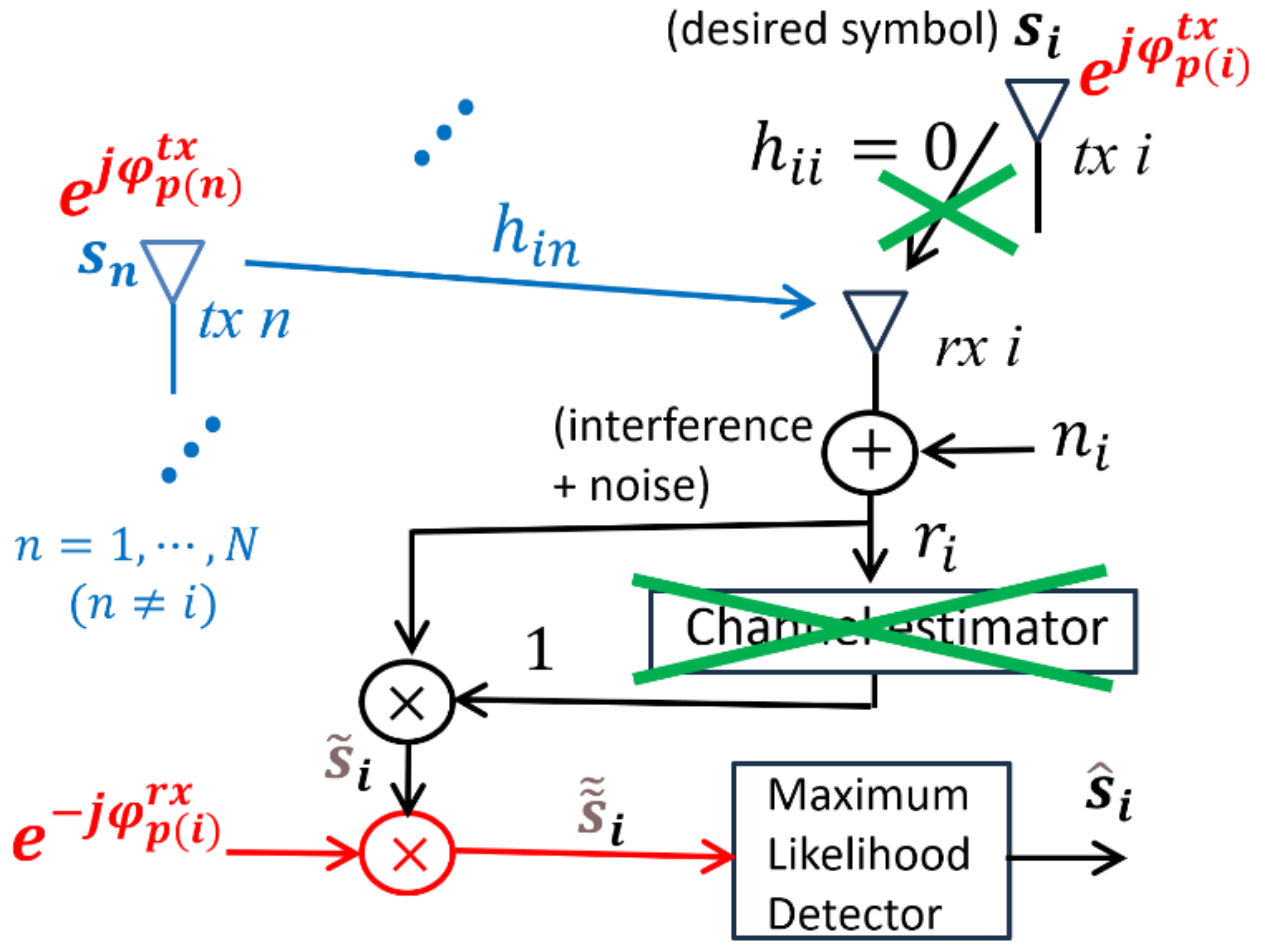

**FIGURE 10 The proposed "though experiment" where own signal is zero because $h_{ii} = 0$ for each co-channel tx.**

Regarding this thought experiment, we pose the following two questions raised in Section I:

**Question 1)** Can the desired transmitted symbol be recovered at the receiver?

**Answer 1)** In the traditional way of thinking, no. This is a "no-hope" case: the desired signal at receiver $i$ is zero because $h_{ii} = 0$, so receiver $i$ obtains nothing but co-channel interference plus thermal noise.

**Question 2)** Can a receiver-side phase, applied before the MLD, help recover the desired symbol?

**Answer 2)** In the traditional way of thinking, no. Since the own signal at receiver $i$ is zero, rotating the co-channel interference (plus noise) at the input of the MLD cannot make a zero signal appear.

We now show that the proposed surrogate-symbol framework of Section III. overturns this "traditional way of thinking."

### B. WHY CANNOT INTERFERENCE CANCELLATION HELP HERE?

In the fully-blocked case in Fig.11, the received signal at rx $i$ (from (22) with $h_{ii} = 0$) is

$$\boldsymbol{r}_i = \boldsymbol{I}_i^{sur} + \boldsymbol{n}_i \tag{55}$$

with no desired-signal term present at all. Interference Cancellation, by construction, estimates and removes $\boldsymbol{I}_i^{sur}$ from $\boldsymbol{r}_i$. Applying it here yields

$$\boldsymbol{r}_i - \boldsymbol{I}_i^{sur} = \boldsymbol{n}_i \tag{56}$$

i.e., pure thermal noise. Interference cancellation is a strategy for *isolating* a desired signal that is already present but obscured by interference; it has nothing to isolate when the desired signal was never there to begin with. The fully-blocked case is therefore precisely the case in which classical interference cancellation is guaranteed to fail, regardless of CSI quality — which makes it the sharpest possible setting in which to test whether the proposed surrogate-symbol approach offers something fundamentally different, rather than a marginal refinement of cancellation-based ideas.

### C. THE PROPOSED SYSTEM UNDER A FULLY-BLOCKED OWN CHANNEL ($h_{ii} = 0$)

The proposed surrogate-symbol-based system for the fully-blocked case is shown in FIGURE 11, under the same Assumptions A1–A5 of Section III. -C (in particular, A4–A5 still require the rx-phase index to be conveyed over a separate, CCI-free control channel within the channel coherence time; this requirement is unaffected by $h_{ii} = 0$, since it concerns only the phase-index side channel, not the blocked direct link). The only structural difference from FIGURE 5 is that the channel-estimator normalization $\boldsymbol{h}_{ii}^{*}/\alpha_{ii}$ is undefined when $\alpha_{ii} = 0$; we therefore replace it with unity gain, which is equivalent to formally setting $\theta_{ii} = 0$ in the derivations of Section IV. (there being no phase reference left to estimate).

With this convention, $\boldsymbol{r}_i$ in (55) is passed directly to the rx-phase multiplication stage, giving

$$\tilde{\tilde{\boldsymbol{s}}}_i = e^{-j\varphi_{p(i)}^{rx}}\,\boldsymbol{r}_i = e^{-j\varphi_{p(i)}^{rx}}(\boldsymbol{I}_i^{sur} + \boldsymbol{n}_i) \quad (57)$$

which is exactly the general result (25) with the own-link magnitude and phase set to zero ($\alpha_{ii} = 0$ and $\theta_{ii} = 0$). In other words, **no new derivation is required**: The fully-blocked case is a literal special case of Proposition 1, obtained simply by setting $\alpha_{ii} = 0$ and $\theta_{ii} = 0$.

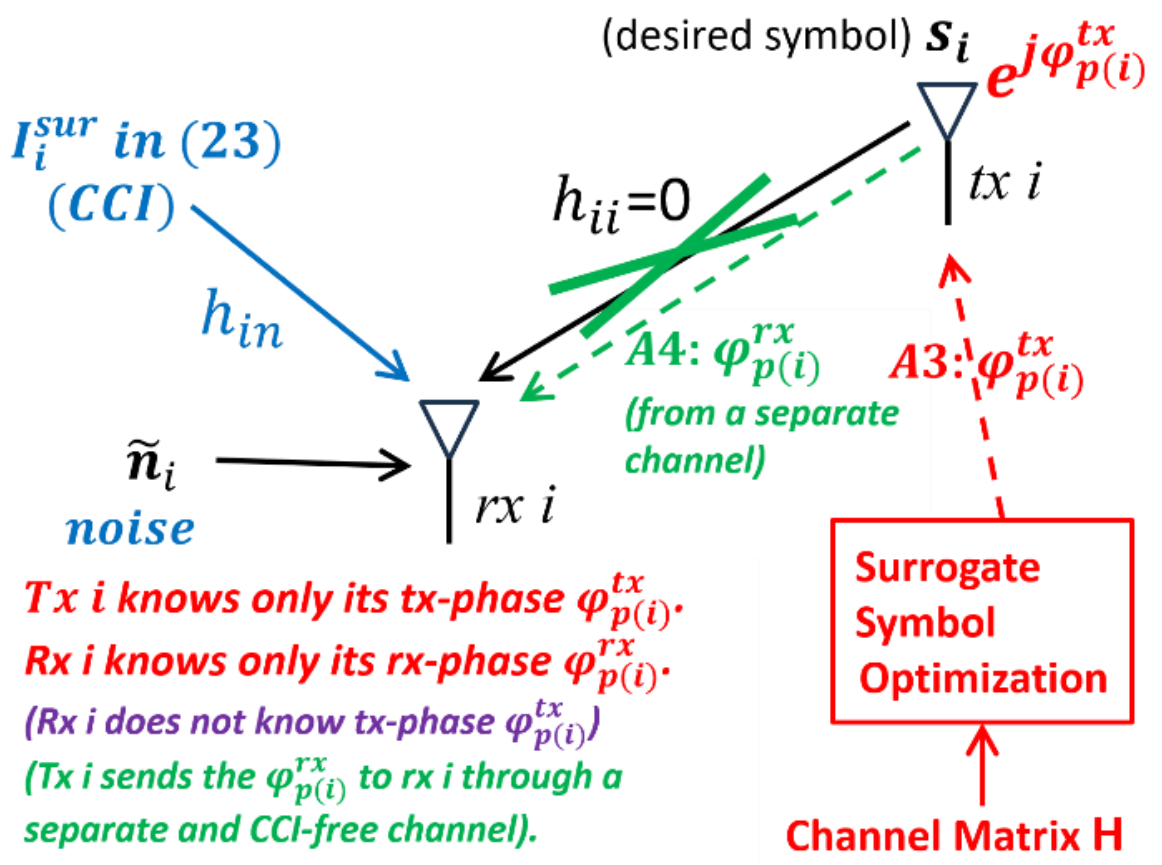

**FIGURE 11 The proposed method when own channel is fully blocked (i.e. $h_{ii} = 0$) in the eyes of the rx *i*.**

### D. RECOVERING THE DESIRED SYMBOL VIA INTERFERENCE SUPPORT

Applying the Interference Support decomposition of Section IV. -B (eq. (46)-(48)) to (57), and using $\alpha_{ii} = 0$, the signal entering the MLD becomes

$$\tilde{\tilde{\boldsymbol{s}}}_i = \beta_i \boldsymbol{s}_i + \boldsymbol{I}_i^{other} + \tilde{\boldsymbol{n}}_i \quad (58)$$

where $\beta_i$ is the magnitude of the Interference Support vector, i.e., $\boldsymbol{I}_i^{sup} = \beta_i\,\boldsymbol{s}_i$, exactly as defined in (48). This equation is the formal resolution of the thought experiment: even though the true desired-signal term is identically zero, the signal entering the MLD still contains a term $\beta_i \boldsymbol{s}_i$ pointing in the direction of the desired symbol $\boldsymbol{s}_i$ — supplied entirely by the co-channel interference itself, reshaped by MaPRIS' phase choice.

This immediately gives:

***Proposition 6.*** *Consider the proposed surrogate-symbol based wireless system in* FIGURE 4 *using the MaPRIS algorithm in* Table 4, *under the fully-blocked condition that every own-channel link is zero. If*

$$d^2\left((\boldsymbol{I}_i^{sup} + \tilde{\boldsymbol{n}}_i),\, \boldsymbol{s}_i\right) \le d^2\left((\boldsymbol{I}_i^{sup} + \tilde{\boldsymbol{n}}_i),\, \boldsymbol{s}_k\right) \quad (59)$$
$$\forall i \ne k, \qquad \forall i = 1, \dots, N$$

then the MLD gives $\hat{\boldsymbol{s}}_1 = \boldsymbol{s}_i$, i.e., the MLD correctly recovers the desired symbol at every receiver simultaneously, despite every own-channel link being zero (i.e. $\boldsymbol{h}_{ii} = 0$).

***Proof:*** This is the network-wide instance of Proposition 4 (Section IV. -B). Crucially, the per-receiver derivation in Section IV. -B — culminating in (58) — depends only on receiver *i*'s own channel row and its own surrogate-symbol optimization; it does not depend on whether any other receiver has a nonzero own-channel link or not. Applying Proposition 4 and the MLD in eq.(7) independently to each receiver under the fully-blocked condition therefore yields (59) simultaneously at every receiver, which completes the proof. ∎

To make the recovery condition (59) quantitative, we define the *effective post-processing SINR* at receiver *i* under the fully-blocked condition as

$$SINR_i^{(h_{ii}=0)} = \frac{E\{\beta_i^{\,2}\}}{E\left\{\left|\boldsymbol{I}_i^{\,other}\right|^2\right\} + E\{|\tilde{\boldsymbol{n}}_i|^2\}} \quad (60)$$

Unlike the general case of Proposition 5, where the own-link gain contributes an interference-independent floor to the numerator, here the entire numerator is supplied by $\beta_i$ — a quantity that exists only because CCI is present and only to the extent that MaPRIS has aligned it with desired symbol $s_i$. This yields a striking corollary:

***Corollary 4.*** *In the fully-blocked case, the presence of CCI is not incidental to recovery — it is the sole source of recoverable signal energy at the receiver. If there is no CCI at all, then* (58) *reduces to pure noise, and no algorithm — MaPRIS included — can recover the desired symbol.*

This sharpens, rather than contradicts, the counter-intuitive claim of Proposition 5: In the general case, higher CCI variance improves SINR on top of an already-nonzero desired link. In the fully-blocked case, CCI is not merely helpful but strictly necessary — it is the only thing standing between the receiver and total blackout.

### E. DISCUSSION: A PROBABILISTIC, NOT GUARANTEED, RECOVERY

Condition (59) is a per-realization condition on the effective angle, the interference norm, and the noise realization; it is not guaranteed to hold for every channel draw and every symbol. Correct recovery becomes more likely as $\beta_i$ grows relative to the residual interference and the noise floor — i.e., as MaPRIS drives the effective angle toward zero in the presence of CCI. In this paper, we do not derive a closed-form recovery probability here, since it depends on the joint distribution of the effective angle and the interference norm over the fading channel and the discrete phase set, which is analytically intractable for general $K$ and $N$. Instead, Section VI. quantifies this probability empirically. There we show that, even in this worst-case scenario, MaPRIS reduces the symbol error rate by roughly an order of magnitude relative to the traditional scheme, which has no mechanism for recovery at all and can only perform at the level of a random guess.

### F. ANALOGY WITH QUANTUM TELEPORTATION

The fully-blocked case has a structural counterpart in the textbook Alice–Bob quantum teleportation example [30], [31], which can be read as a lock-and-key story. Alice wants to send an unknown qubit state to Bob. What Bob holds is a locked box: his qubit contains the state, but scrambled by one of four random Pauli operations, so taken alone it is independent of the state and carries no information about it (a quantum one-time pad). The key is Alice's two classical bits, sent to Bob over an ordinary classical channel (a phone call would do). They too look random on their own. Bob uses the key to select and apply the matching rotation (the Pauli correction), the box opens, and the state appears: the information resides neither in the box nor in the key, but in their combination. In the proposed surrogate-symbol system with $\boldsymbol{h}_{ii} = 0$, the same holds. The received signal $r_i$ contains interference only and, taken alone, is independent of the desired symbol $s_i$: it is the locked box. The key is the rx-phase index sent over the separate control channel (Assumption A4); it looks uninformative to the receiver, because the receiver does not know the transmit phase $\varphi^{tx}_{p(i)}$. Turning the key means applying the rotation $e^{-j\varphi^{rx}_{p(i)}}$ selected by that index, which turns the interference into desired symbol $s_i$; the MLD then reads out what is inside. The role of the pre-shared entangled pair, which makes the box openable at all, is played here by the phase optimization, which aligns the interference with the surrogate symbol (Section IV-B). Table 6 sets the two side by side.

**Table 6 Quantum teleportation (Alice–Bob) versus surrogate symbols.**

| Teleportation | Surrogate symbols (this paper) |
|---|---|
| Entanglement: pre-established entangled pair | Correlation: A correlation between the surrogate interference and the surrogate symbol is created (by the phase optimization – game of phases) |
| Bob's qubit: **the locked box** (uninformative alone) | $r_i$ when $h_{ii} = 0$: **the locked box**, interference only, independent of $s_i$ |
| Two classical bits: **the key**, sent over a classical channel (uninformative alone) | The phase index on the control channel: **the key** (uninformative alone) |
| **Turning the key**: the Pauli correction selected by the bits | **Turning the key**: the rotation $e^{-j\varphi^{rx}_{p(i)}}$ |
| Reading what is inside the opened box | The MLD |

The analogy has limits. (i) In teleportation the key is uniformly random and independent of the state, which makes the box perfectly secure; here, for a given channel matrix $H$, the key is a function of the desired symbol, and it appears uninformative only because the receiver does not know the transmit phase. This is a similarity, not an equivalence, and no secrecy property is claimed for the proposed system. (ii) In teleportation the resource is a pre-established entangled pair; here the corresponding resource is created by the phase optimization: by minimizing the effective angle, MaPRIS aligns the interference with the surrogate symbol, producing a classical correlation (phase alignment) that the key then converts into the desired symbol. The optimization plays the role that the entangled pair plays, but the correlation is classical, not entanglement. (iii) Teleportation recovers the state exactly after the correction, whereas recovery here is probabilistic (see the preceding discussion). (iv) The proposed system is entirely classical; the analogy is structural and carries no claim about quantum physics.

## VI. SIMULATION RESULTS

To the best of our knowledge, the proposed surrogate-symbol based co-channel wireless system in FIGURE **4** as well as the "game of phases" optimization problem in FIGURE **6** are both novel, and therefore there are no reference algorithms other than the optimum standard wireless system in **FIGURE 2.** That's why we consult the complex graph and complex neural networks literature and borrow the following algorithms for the simulation results just for comparison reasons.

### A. COMPLEX HOPFIELD NEURAL NETWORK (CHNN) BASED SURROGATE SYMBOL DETERMINATION

The cost function of the CHNN algorithm is given by

$$J_{CHNN}\left(\mathbf{H}, \{\varphi^{tx}_{p(i)}\}_{i=1}^{N}\right) = -\sum_{n=1}^{N} e^{-j\varphi^{tx}_{p(i)}} \boldsymbol{I}_{\boldsymbol{n}}^{ph}\left(\{\varphi^{tx}_{p(i)}\}_{i=1}^{N}\right) = -\sum_{n=1}^{N} e^{-j\varphi^{tx}_{p(n)}} \sum_{\substack{i=1 \\ (i\neq n)}}^{N} \alpha_{ni} e^{j\left(\varphi^{tx}_{p(i)} - \theta_{ni}\right)} \tag{61}$$

This cost function follows the standard Hopfield-network energy formulation [11], adapted here to the complex-valued,

phase-based setting of our problem. The surrogate symbols are determined by the CHNN optimization as follows:

$$\arg_{\{\varphi_{p(i)}^{tx}\}_{i=1}^{N}} \min J_{CHNN}\left(\mathbf{H},\{\varphi_{p(i)}^{tx}\}_{i=1}^{N}\right) \quad (62)$$

### B. SHADOW-CUT [2] BASED SURROGATE SYMBOL DETERMINATION

The cost function of the Shadow-Cut phase optimization algorithm in [2] is defined as follows:

$$J_{SC}\left(\mathbf{H},\{\varphi_{p(i)}^{tx}\}_{i=1}^{N}\right) = -\sum_{k=1}^{K}\left(\sum_{m\in C_k}\sum_{z\in \bar{C}_k} \alpha_{mz} e^{j\left(\theta_{mz}+\varphi_{p(m)}^{tx}-\varphi_{p(z)}^{tx}\right)}\right) \quad (63)$$

where $C_k$ and $\bar{C}_k$ represents the set of indices of those transmitters (neurons) in phase $k$, and the set of the indices of the rest of the transmitters (neurons) which are not in phase $k$. The surrogate symbols are determined by the SIGA algorithm in [2] as follows:

$$\arg_{\{\varphi_{p(i)}^{tx}\}_{i=1}^{N}} \min J_{SC}\left(\mathbf{H},\{\varphi_{p(i)}^{tx}\}_{i=1}^{N}\right) \quad (64)$$

Without loss of generality and for the sake of brevity, we present the results of the standard wireless system in **FIGURE 2** and the proposed surrogate-symbol based wireless system in **FIGURE 4** which contains 8 tx-rx pairs and for 4-PSK case only. In all simulations, there are 8 surrogate symbols (phases):

$$P = \{e^{-j\varphi_k}\}_{k=1}^{k=8} = \left\{e^{-j0}, e^{-j\frac{\pi}{4}}, e^{i\frac{\pi}{4}}, e^{-j\frac{\pi}{2}}, e^{j\frac{\pi}{2}}, e^{-j\frac{3\pi}{4}}, e^{j\frac{3\pi}{4}}, e^{j\pi}\right\} \quad (65)$$

So, the surrogate symbols $\boldsymbol{s}_i^{sur} \in P$, where $i=1,2,...,8$. We conduct all our experiments using MATLAB R2024b. The wireless system simulated is shown in **FIGURE 4** with 8 tx-rx pairs all of which are using the same channel (i.e. co-channel). The baseband channel coefficient $\boldsymbol{h}_{ij}$ is Rayleigh distributed. Extensive simulation studies have been conducted, and the results validate the findings presented in this paper.

In order to give an insight into the evolutions of the proposed algorithms MiPIP and MaPRIS, we present their cost functions' evolutions for some random snapshots of the system in Fig.4. Later on, we present the results of our simulation campaigns for two different scenarios: medium CCI case and high CCI case.

#### 1) SIMULATION SETUP

The simulated network consists of $N = 8$ single-antenna co-channel tx-rx pairs sharing the same time-frequency resource as shown in **FIGURE 4**. Each transmitter sends a Gray-coded QPSK symbol (4-PSK, phase offset π/4), so all transmitted symbols have unit magnitude.

The link-gain matrix G is generated as flat Rayleigh fading: each entry $\boldsymbol{g}_{ij} = X + jY$, with X, Y ~ $\mathcal{N}(0, 0.5)$ drawn independently and identically for every *(i,j)* pair, including the desired-link *(i=j)* entries. This circularly symmetric complex Gaussian construction gives unit average link power for every link in the matrix (i.e. $\Omega = E\left\{\left|\boldsymbol{h}_{ij}\right|^2\right\} = 1$). To control the CCI level independently of the desired-link gain, the raw channel matrix is scaled entry-wise, $\boldsymbol{h}_{ij} = b_{ij}\boldsymbol{g}_{ij}$, where the weighting coefficient $b_{ij}$ depends on the role of link *(i,j)*: For desired-link $b_{ij} = b_1 = 1$ (for $i = j$), and for CCI link $b_{ij} = b_2 = 10^{-E_2/10}$ for CCI link (for $i \neq j$). So, the desired-link weight is held fixed at $E = 0\ dB$ in all scenarios, so only the interference level is varied and any change in link performance is attributable solely to CCI.

Each receiver is corrupted by additive complex Gaussian noise with total power $P_{noise} = P_{data} \cdot 10^{-SNR/10}$, with $P_{data} = \sqrt{2}$ and SNR = 40 dB fixed across all scenarios, giving $P_{noise} = 1.4142 \times 10^{-4}$. This value is deliberately kept low and identical across scenarios so the system operates in the interference-limited regime, isolating the effect of CCI from thermal noise.

For a given receiver, the average desired-signal power is $P_d$ and the average aggregate CCI power $P_I$ from the remaining $N-1 = 7$ co-channel tx's is

$$P_d = b_1^2\Omega = 1^2 1 = 1 \quad (66)$$
$$P_I = (N-1)b_2^2\Omega = (8-1)b_2^2 1 = 7\, b_2^2$$

The SIR, SNR and SINR is obtained as follows

$$SIR\ (\text{dB}) = 10\log 10\left(\frac{P_d}{P_I}\right) \quad (67)$$
$$SNR\ (\text{dB}) = 10\log 10\left(\frac{P_d}{P_{noise}}\right)$$
$$SINR\ (\text{dB}) = 10\log 10\left(\frac{P_d}{P_I + P_{noise}}\right)$$

$P_{noise}$ is normalized quantity, just like $P_{data}$, **H**, and the desired and interference powers $P_d$ and $P_I$ computed from it. This is standard practice in baseband link-level simulations: Everything is defined relative to the unit-magnitude PSK symbol, so $P_{noise}$, $P_d$, and $P_I$ are all in the same arbitrary "normalized power" units. What's physically meaningful is only the *ratios* between them (SIR, SNR, SINR in dB), not their individual absolute values.

Using the interference-weighting parameters for the Medium- and High-CCI scenarios, Table 7 lists the resulting average SIR, SNR, and SINR, computed from $\Omega = 1$ and $N = 8$. Because thermal noise is roughly five to six orders of magnitude weaker than the interference power in all scenarios, SINR ≈ SIR (i.e. CCI-limited scenarios).

**Table 7 Interference-weighting parameters and resulting SNR and SINR.**

| Scenario | Desired power $P_d$ | CCI power $P_I$ | SNR (dB) | **SIR≈SINR (dB)** |
|---|---|---|---|---|
| Medium CCI | 1.00 | 27.87 | 38.49 | **-14.45** |
| High CCI | 1.00 | 278.68 | 38.49 | **-24.45** |

### 2) MAPRIS COST FUNCTION EVOLUTION

In this section, we give an insight into how the MaPRIS cost function $J_{MaPIS}$ in (34) evolves by time. A typical evolution is shown in FIGURE 12 for a random snapshot in Fig.4 for our 8 co-channel transmitters case. As seen from the figure, the MaPRIS strictly increases for each phase update (i.e., surrogate symbol update) and converges after several steps, which confirms Proposition 3 above.

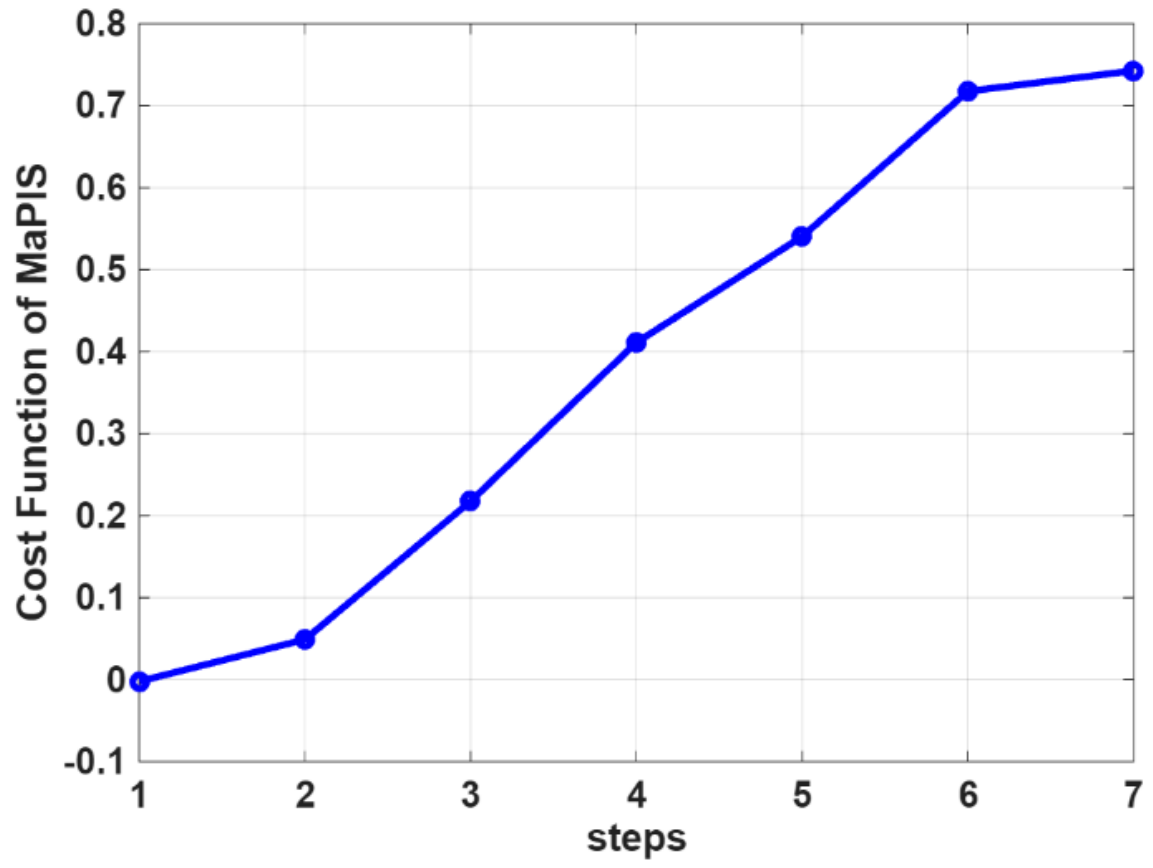

**FIGURE 12. The evolution of the MaPRIS cost function $J_{MaPIS}$ in (34) for a random channel realization of the system in FIGURE 4. (At each step, a surrogate symbol is updated).**

In section IV above, the promise of the proposed MaPRIS in Proposition 3 is to maximize $J_{MaPIS} = E\{cos(\Xi_i^{eff})\}$, or equivalently the average of $|I_i^{sup}|$ as shown in (52), over $N$ tx-rx pairs. How about $E\{|\Xi_i^{eff}|\}$? The FIGURE 13 shows the evolution of $E\{|\Xi_i^{eff}|\}$ of the MaPRIS with $E\{cos(\Xi_i^{eff})\}$ for the same snapshot as in FIGURE 12. As seen from FIGURE 13, the MaPRIS with (34) tends to decrease $E\{|\Xi_i^{eff}|\}$, as expected, although it does not necessarily promise a monotonic decrease because it is a (slightly) different cost function. As explained in Section IV, while the cost function $J_{MaPRIS}$ with (34) maximizes $E\{cos(\Xi_i^{eff})\}$, or equivalently $E\{|I_i^{sup}|\}$ as shown in (52), the MaPRIS with (35) minimizes the $E\{|\Xi_i^{eff}|\}$ as shown in (53). In order to give an insight into the rx-specific effective angle $|\Xi_i^{eff}|$, we present FIGURE 24 in the Appendix showing the initial step and the final step (after convergence) of the MaPRIS for each receiver from rx-1 to rx-8. As seen from FIGURE 24, the *efficient angle* $|\Xi_i^{eff}|$ (i.e., the angle between the *phased and rotated CCI* and the *desired symbol*) is decreased for each rx except rx 5 by the MaPRIS. This is because in our simulations, we used $J_{MaPRIS}$ with (34) which maximizes $E\{cos(\Xi_i^{eff})\}$. FIGURE 12, FIGURE 13 and FIGURE 24 confirm Proposition 3 and Corollary 3.

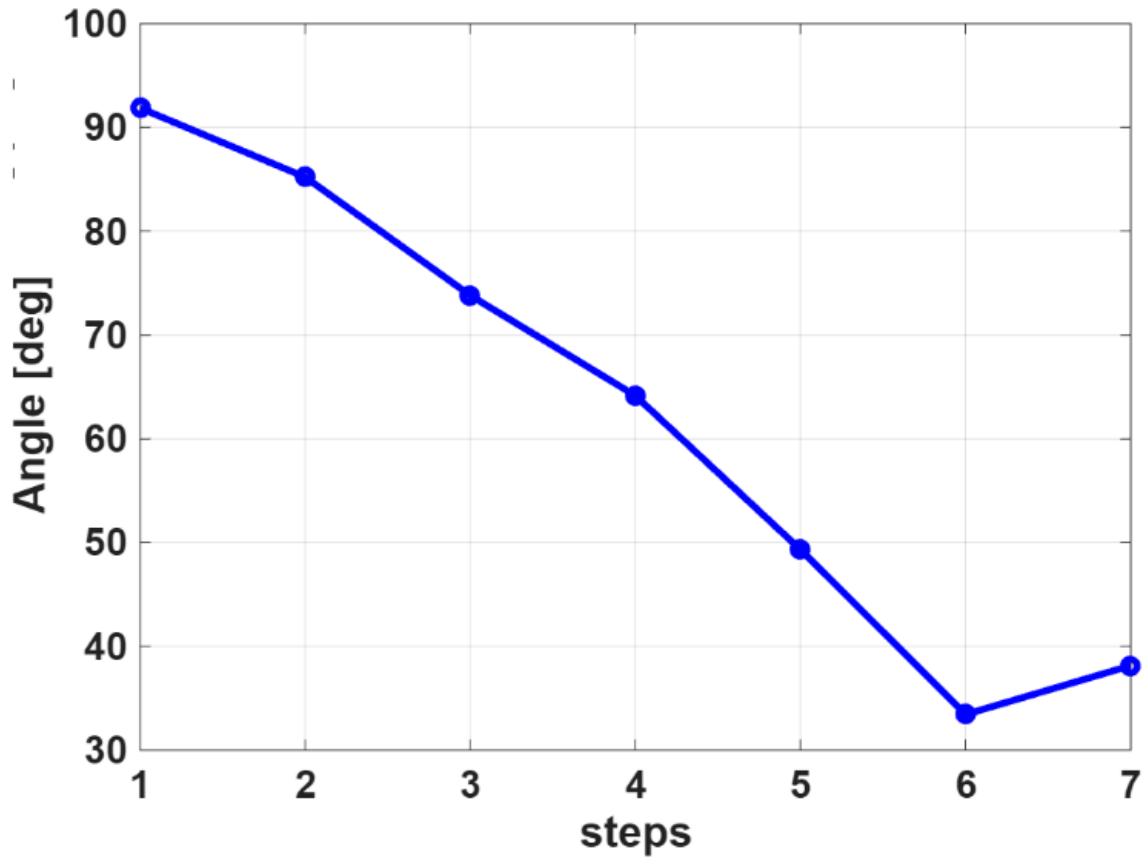

**FIGURE 13. The evolution of the average of the $|\Xi_i^{eff}|$ in MaPRIS with $J_{MaPIS}$ in (34) (at each step, a surrogate symbol is updated).**

### 3) MIPIP COST FUNCTION EVOLUTION

In this section, we give an insight into how the MaPRIS cost function $J_{MiPIP}$ in (29) evolves by time. A typical evolution with respect to the surrogate symbol updates for a random channel realization of the wireless system in FIGURE **4** is shown in **FIGURE 14**. As seen from the figure, the MiPIP strictly decreases for each phase update (i.e., surrogate symbol update) and converges after several steps, which confirms Proposition 2.

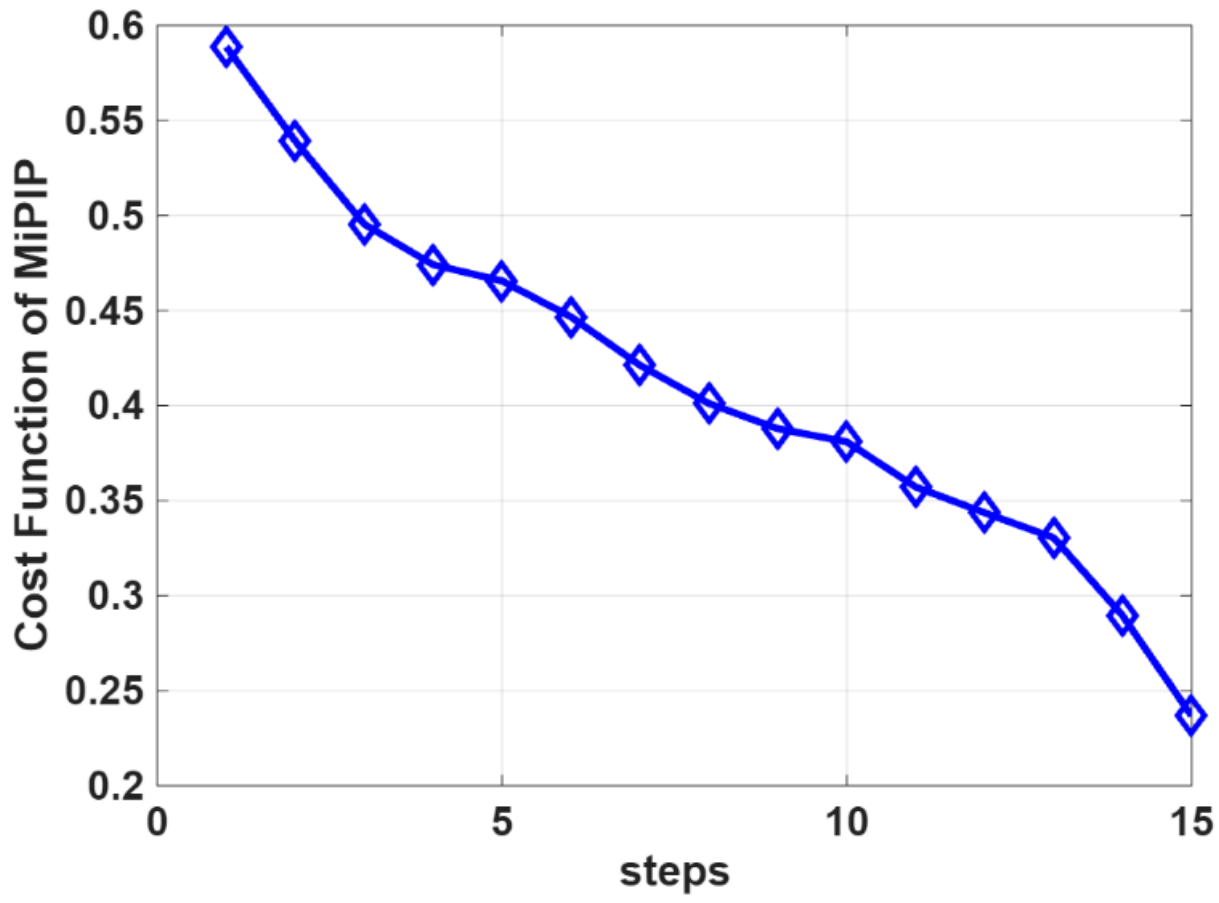

**FIGURE 14. The evolution of the MiPIP cost function $J_{MiPIP}$ in (29) for a random channel realization in FIGURE 4 .**

In order to give an idea about the evolution of the *average effective angle* (i.e., $E\{|\Xi_i^{eff}|\}$) during the MiPIP optimization, we plot it in **FIGURE 15** for the same channel realization. As seen from the figure, the *average effective angle* $|\Xi_i^{eff}|$ randomly changes during the MiPIP optimization, simply because the MiPIP does not take the *effective angle* $\Xi_i^{eff}$ into account at all, which is in line with Proposition 2 and Proposition 3.

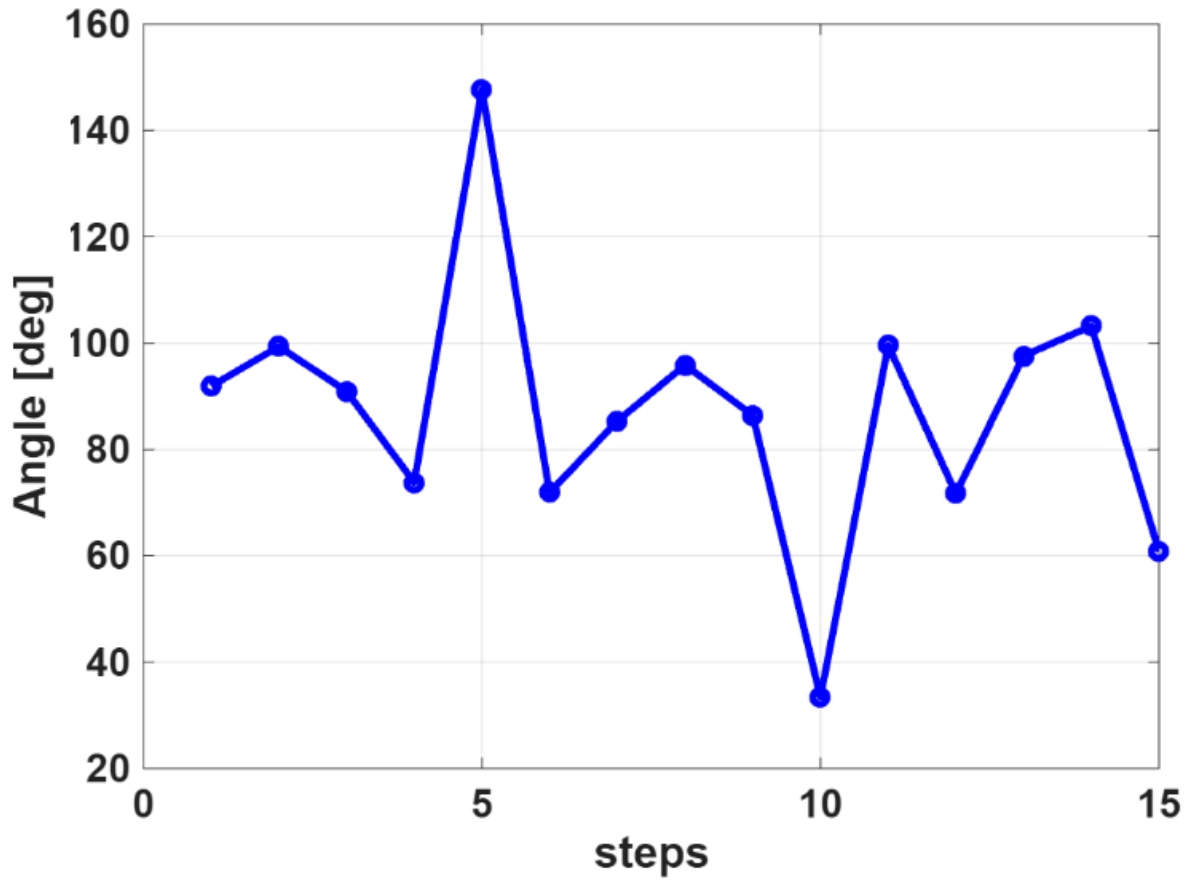

**FIGURE 15. The evolution of the average of the average *effective angle* $|\Xi_i^{eff}|$ in MiPIP (at each step, a surrogate symbol is updated).**

### 4) MEDIUM CCI CASE

In this section, we present the system-level average BER and Symbol Error Rate (SER) performance results for a medium interference scenario of the 8 rx-tx wireless system in FIGURE **4**. The algorithms used in our simulations are listed in Table 8.

The algorithms A, B, C and E are used only for comparison reasons as well as for verification reasons of our developed simulator. We used 4-PSK modulation. For example, algorithm A theoretically is expected to give SER=0.75 and BER=0.5 for 4-PSK modulation. Indeed, our simulation results below confirm this theoretical result. Similarly, algorithm B (the opposite of MaPRIS) and algorithm C (the opposite of MiPIP) are supposed to give worse performance than the rest of the algorithms because they deliberately move interference support away from the desired symbol and deliberately maximize the interference norm, respectively. Indeed, our simulation results below confirm these expectations.

**Table 8.** Algorithms used in the simulations

| Name | Phase algorithm |
|---|---|
| A | Tx phase $\varphi_{p(i)}^{tx}$ and Rx phase ($\varphi_{p(i)}^{rx}$) is random |
| B | Opposite of MaPRIS: *i.e.,* $\min E\{cos(\Xi_i^{eff})\}$ |
| C | Opposite of MiPIP: i*.e.,* $\max E\{\|\boldsymbol{I}_i^{sur}\|\}$ |
| D | SIGA: $\min J_{SIGA}$ in (64) |
| E | Random $\varphi_{p(i)}^{tx} = \varphi_{p(i)}^{rx}$ |
| **F** | **Traditional case** (FIGURE 2) |
| G | CHNN: $\min J_{CHNN}$ in (61) |
| **H** | **MiPIP:** $\min E\{\|\boldsymbol{I}_i^{sur}\|\}$ in (29) |
| **I** | **MaPRIS:** $\max E\{cos(\Xi_i^{eff})\}$ in (34) |

The average SER and average BER results are shown in FIGURE 16.*(a)* and *(b)* respectively. As seen from the figure, the best performance is achieved by the proposed MaPRIS by far, and the 2nd best performance is achieved by the proposed MiPIP.

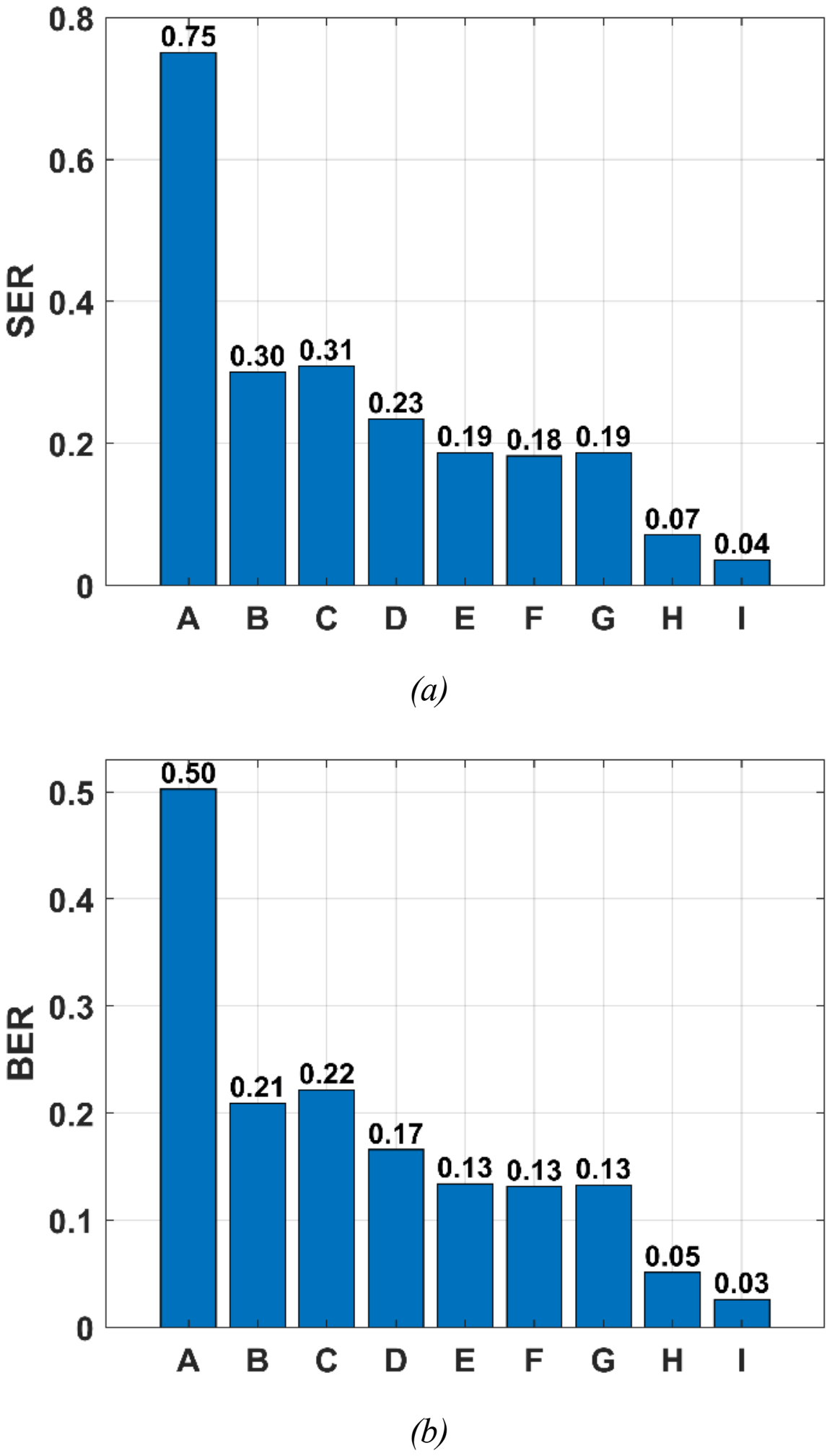

**FIGURE 16. (*a*) Overall average SER results, and (*b*) BER results for the medium CCI case.**

FIGURE 17.*(a)* and *(b)* presents the rx-specific average SERs and BERs, respectively. In the figure, rx *i* represents receiver (mobile) *i*. Both rx-specific SER and BER results in FIGURE 17 are in line with the network-level SER and BER results in FIGURE 16, and show that each user, without exception, enjoys the superior performance of the proposed MaPRIS and MiPIP algorithms. These results also confirm that the algorithms D (i.e., SIGA in (64)) and algorithm G (i.e., CHNN in (61)) does not give superior performance as compared to the traditional wireless system in FIGURE 2 (i.e., algorithm F).

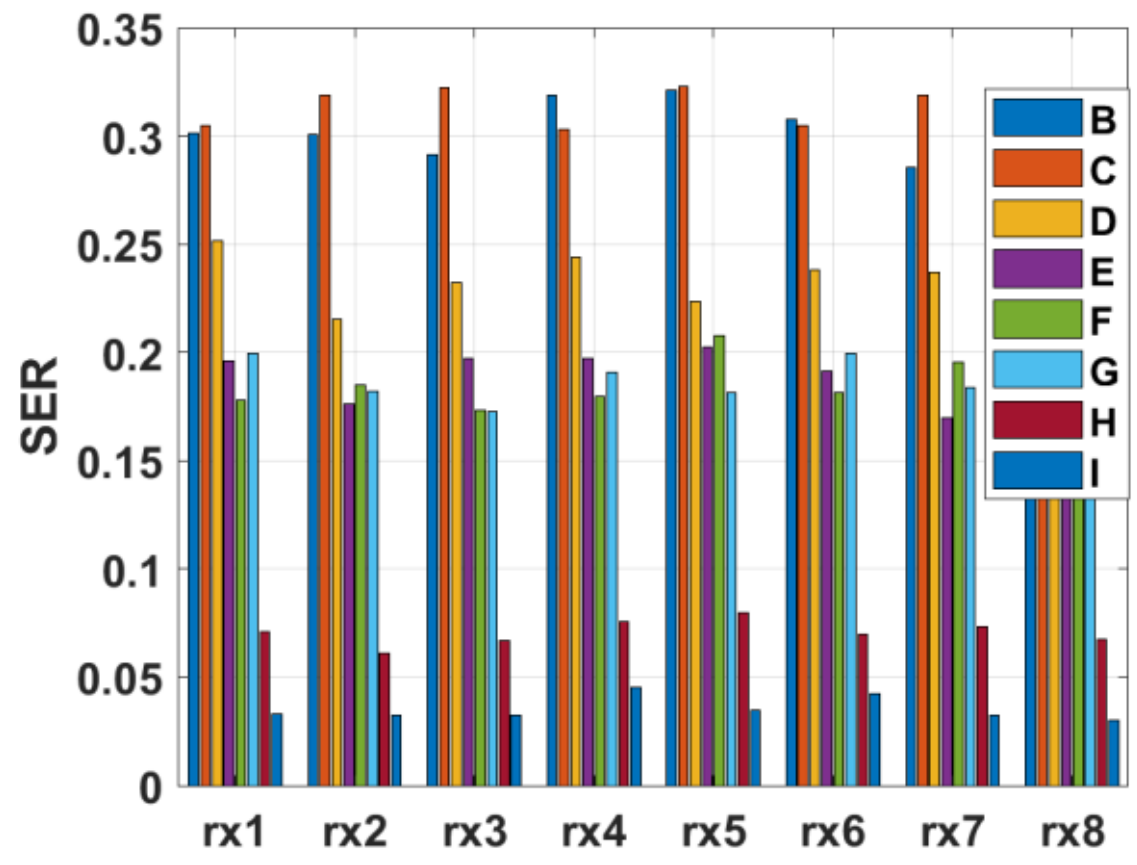

*(a)*

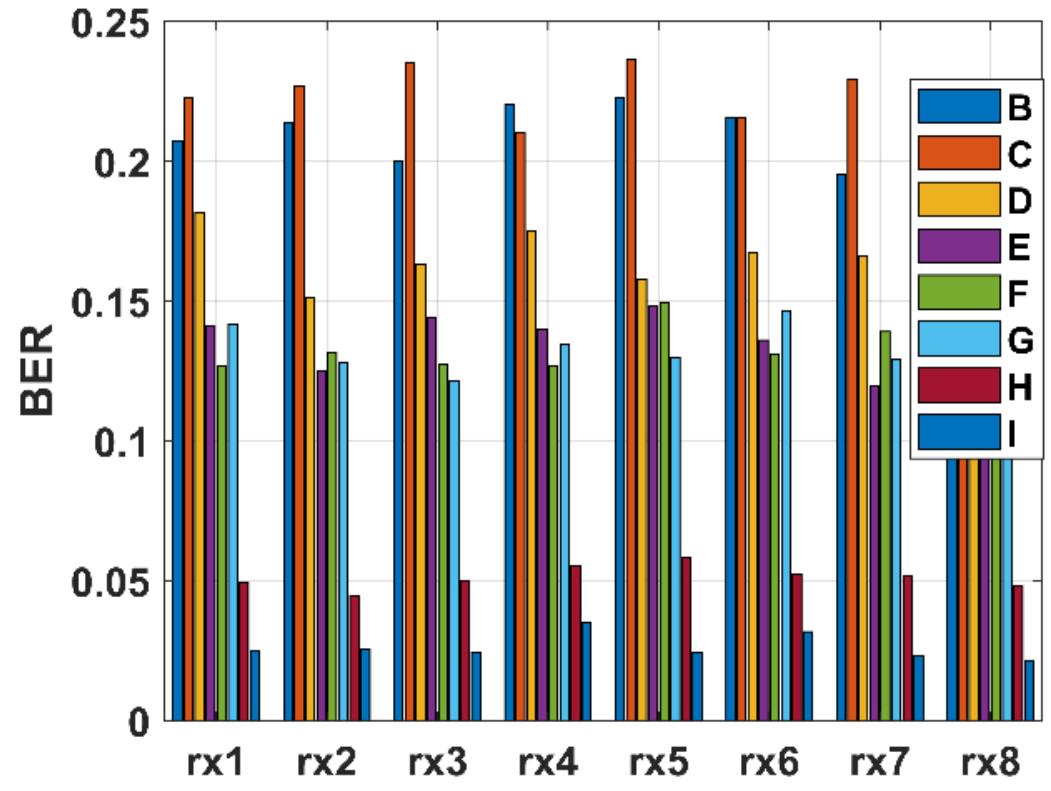

*(b)*

**FIGURE 17. (*a*) Rx-specific SER results, and (*b*) Rx-specific BER results for the medium CCI case.**

CDF of SER and CDF of BER of the traditional system in FIGURE 2 and the proposed methods MaPRIS and MiPIP are presented in FIGURE 18.(*a*) and FIGURE 18.(*b*), respectively (for the medium CCI case). The figures confirm Proposition 2 and 3, i.e., the proposed algorithms MaPRIS and MiPIP give superior performance as compared to the traditional method in FIGURE 2.

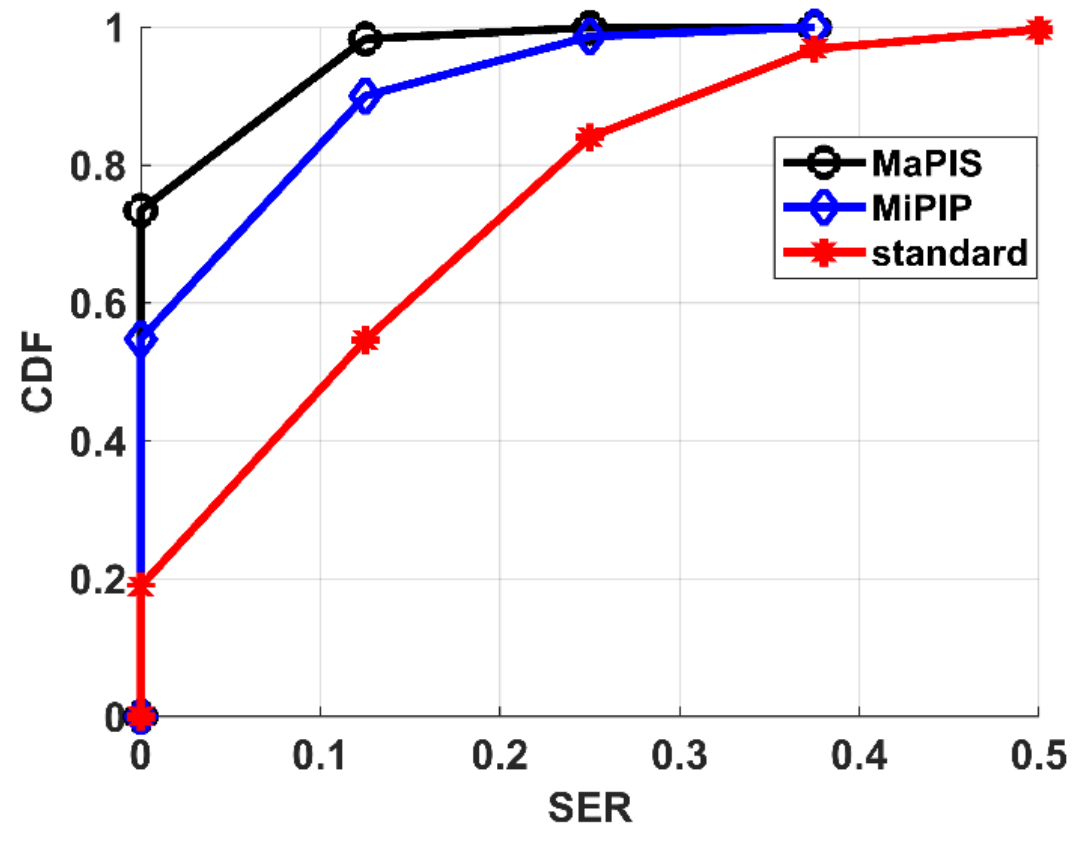

*(a)*

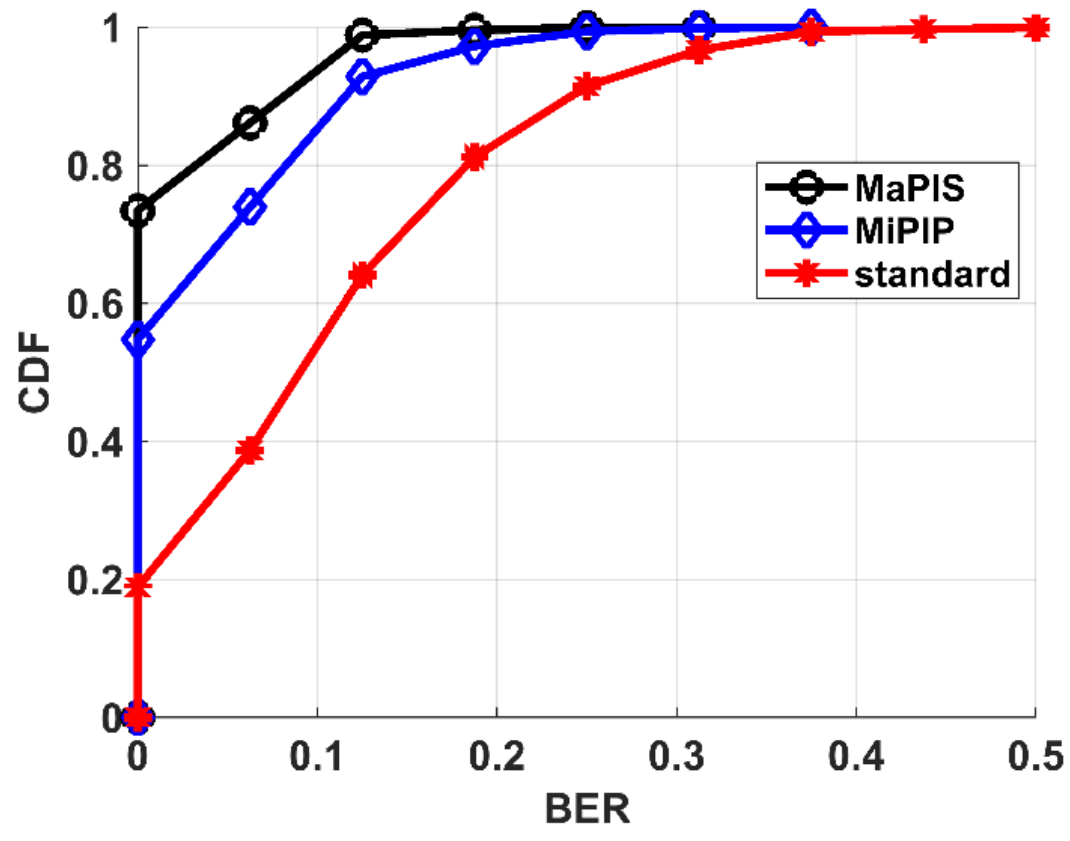

*(b)*

**FIGURE 18. Comparison of (*a*) CDFs of SER and, (*b*) CDFs of BER for the proposed and standard methods in the medium CCI case.**

### 5) HIGH CCI CASE

In this section, we present the system-level average BER and SER performance results for a high CCI scenario of the 8 rx-tx wireless system proposed in FIGURE **4**. The average SER and average BER results are shown in FIGURE 19.*(a)* and *(b)* respectively. As seen from the figure, the best performance is achieved by the proposed MaPRIS by far, and the 2nd best performance is achieved by the proposed MiPIP.

FIGURE 20.*(a)* and *(b)* presents the mobile-specific average SERs and BERs, respectively. In the figure, rx *i* represents receiver (mobile) *i*. Both mobile-specific SER and BER results in FIGURE 17 are in line with the network-level SER and BER results, and show that each user, without exception, enjoys the superior performance of the proposed MaPRIS and MiPIP as compared to the traditional method in FIGURE 2 (algorithm F).

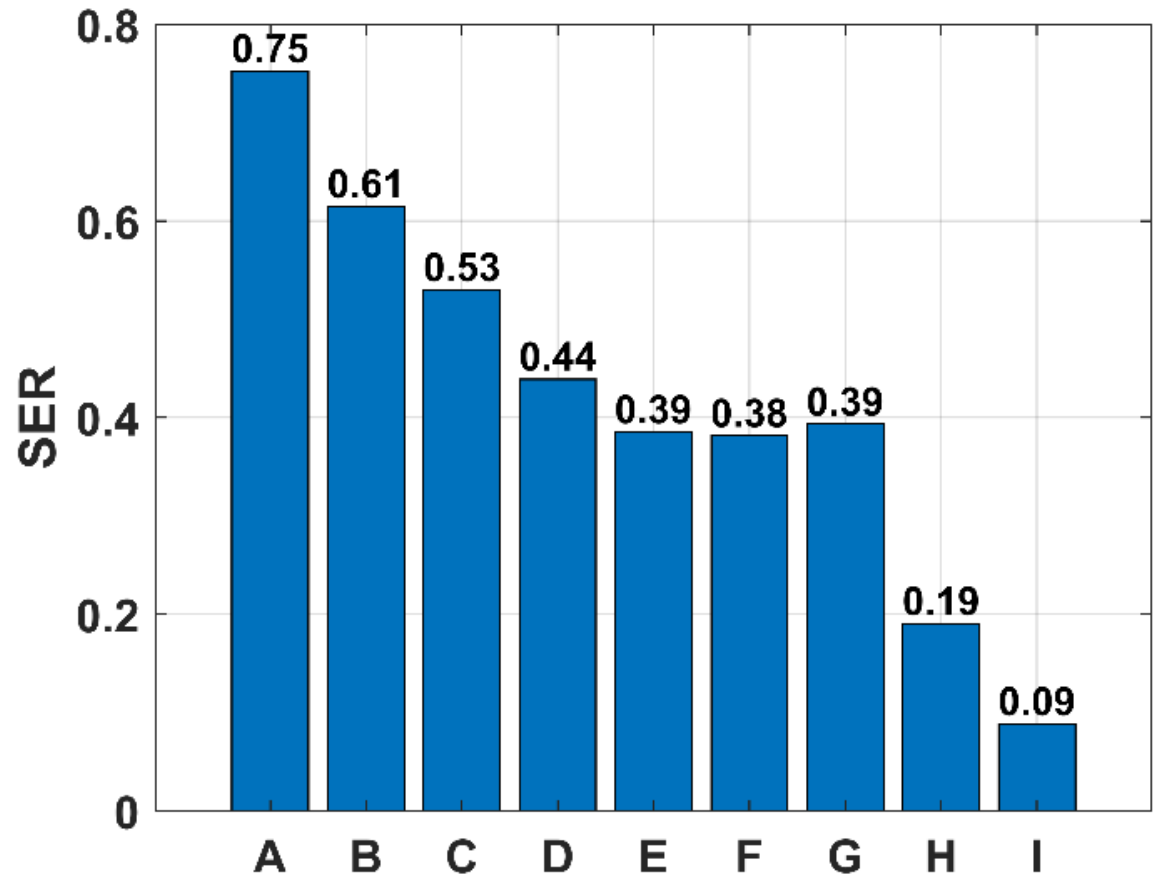

*(a)*

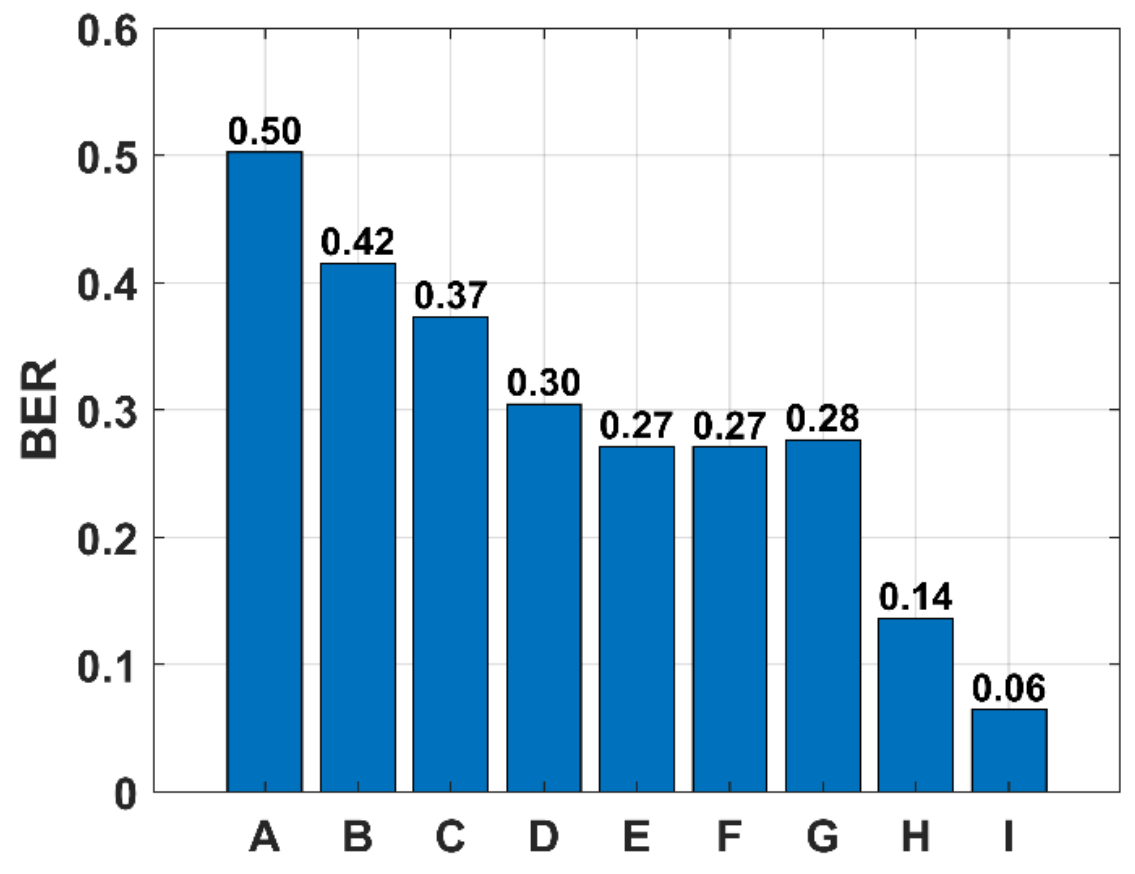

*(b)*

**FIGURE 19. (*a*) Overall average SER results, and (*b*) BER results for the high CCI case.**

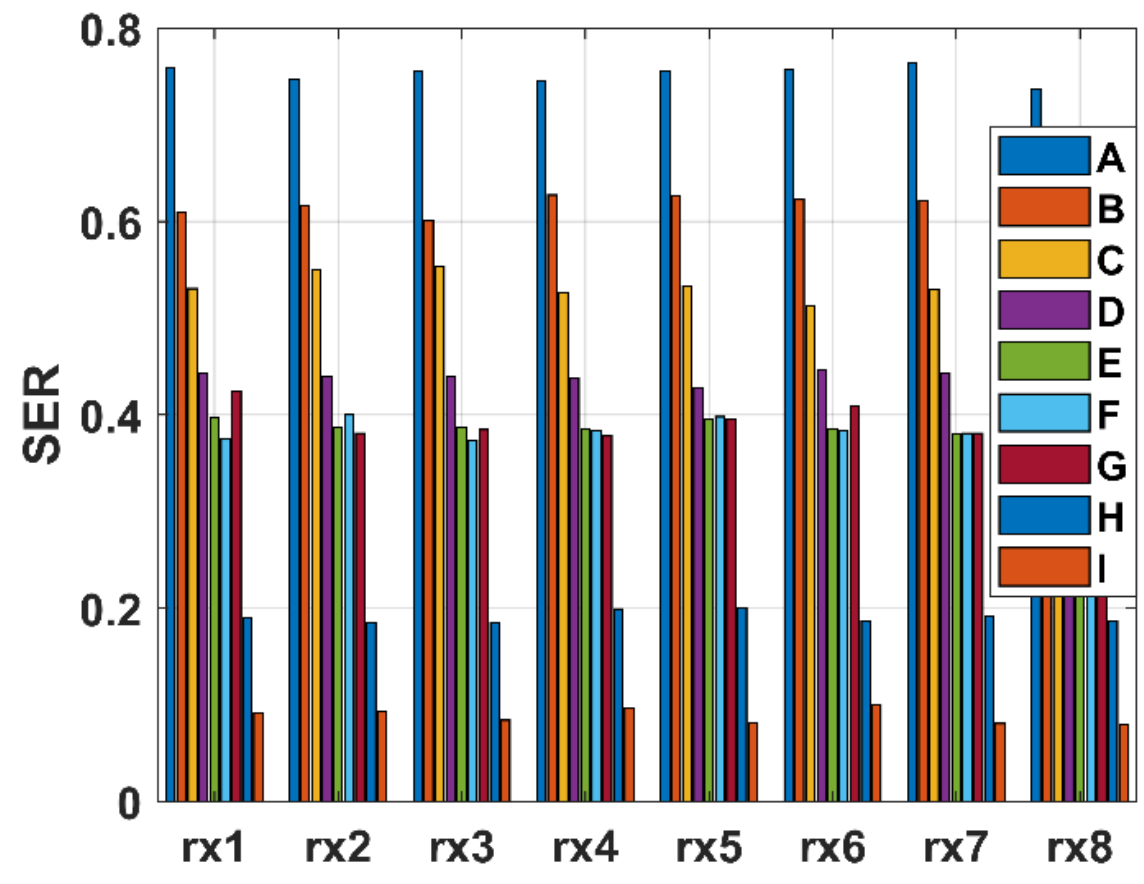

*(a)*

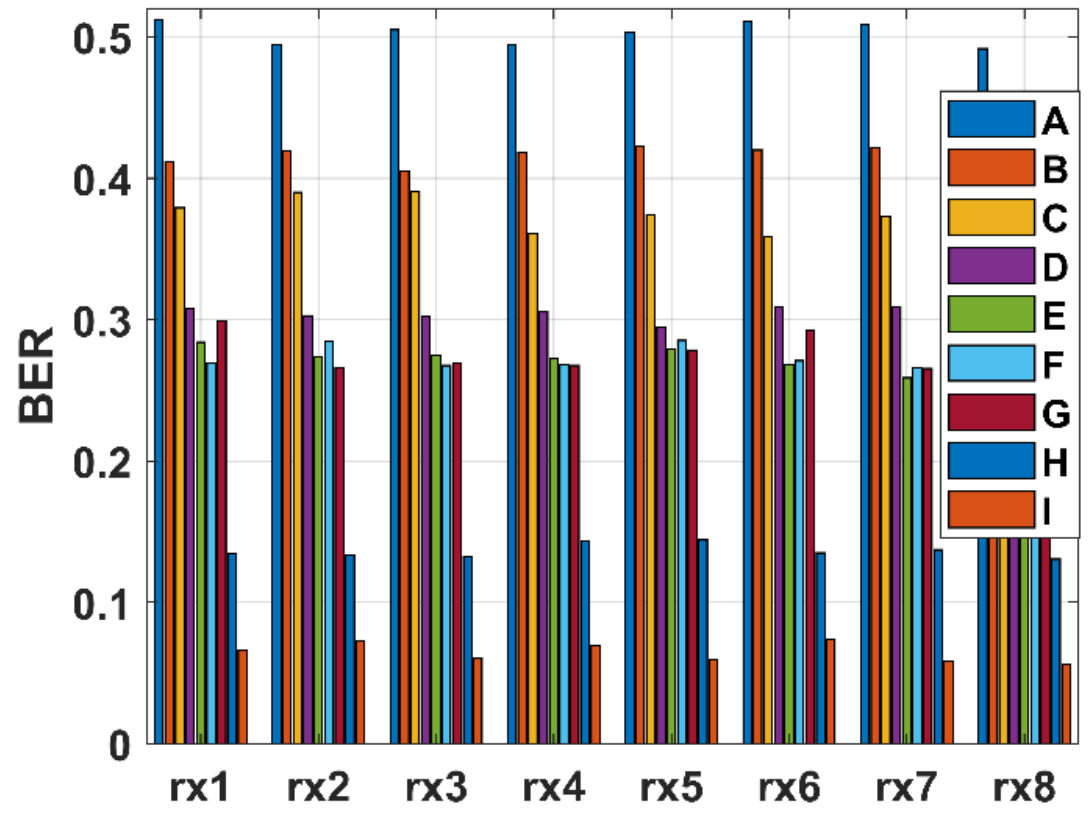

*(b)*

**FIGURE 20. (*a*) Rx-specific SER results, and (*b*) Rx-specific BER results for the high CCI case.**

CDF of SER and CDF of BER of the traditional method and the proposed methods (MaPRIS and MiPIP) in **FIGURE 4** are presented in FIGURE 21.(*a*) and FIGURE 21.(*b*), respectively (for a high CCI case). The figures confirm that the proposed algorithms MaPRIS and MiPIP give superior performance as compared to the standard method (**FIGURE 4**).

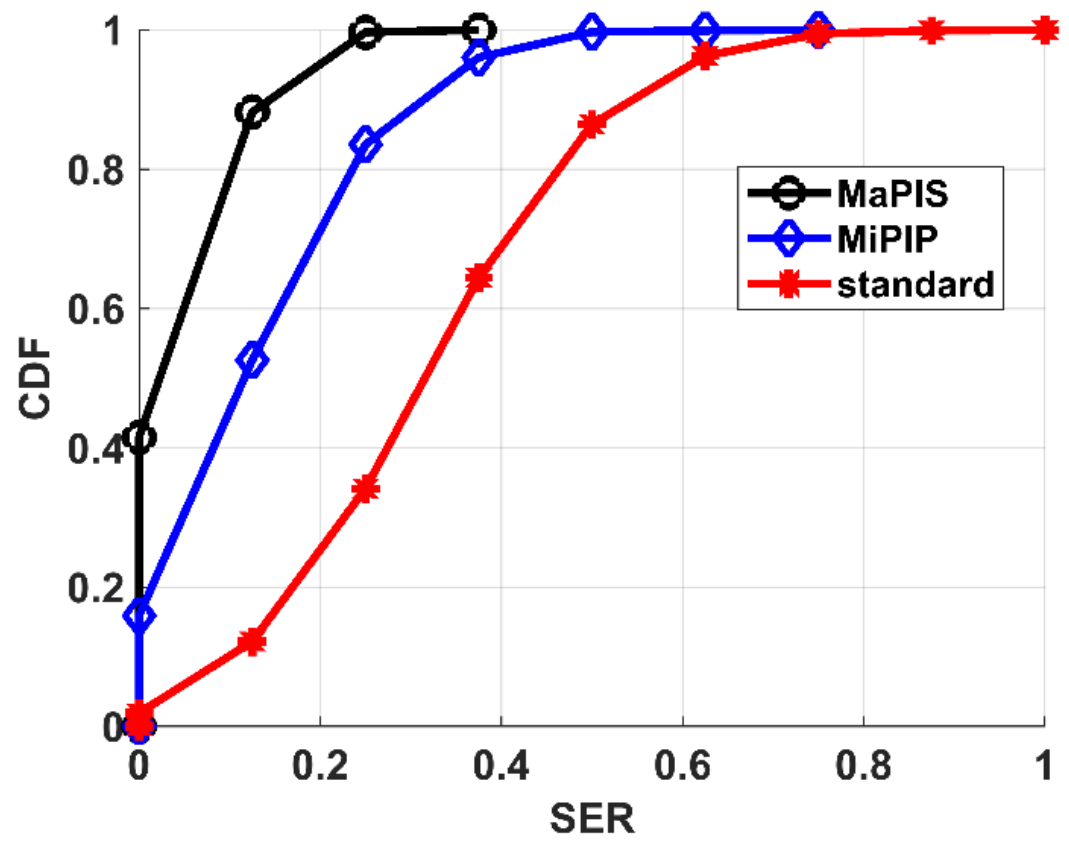

*(a)*

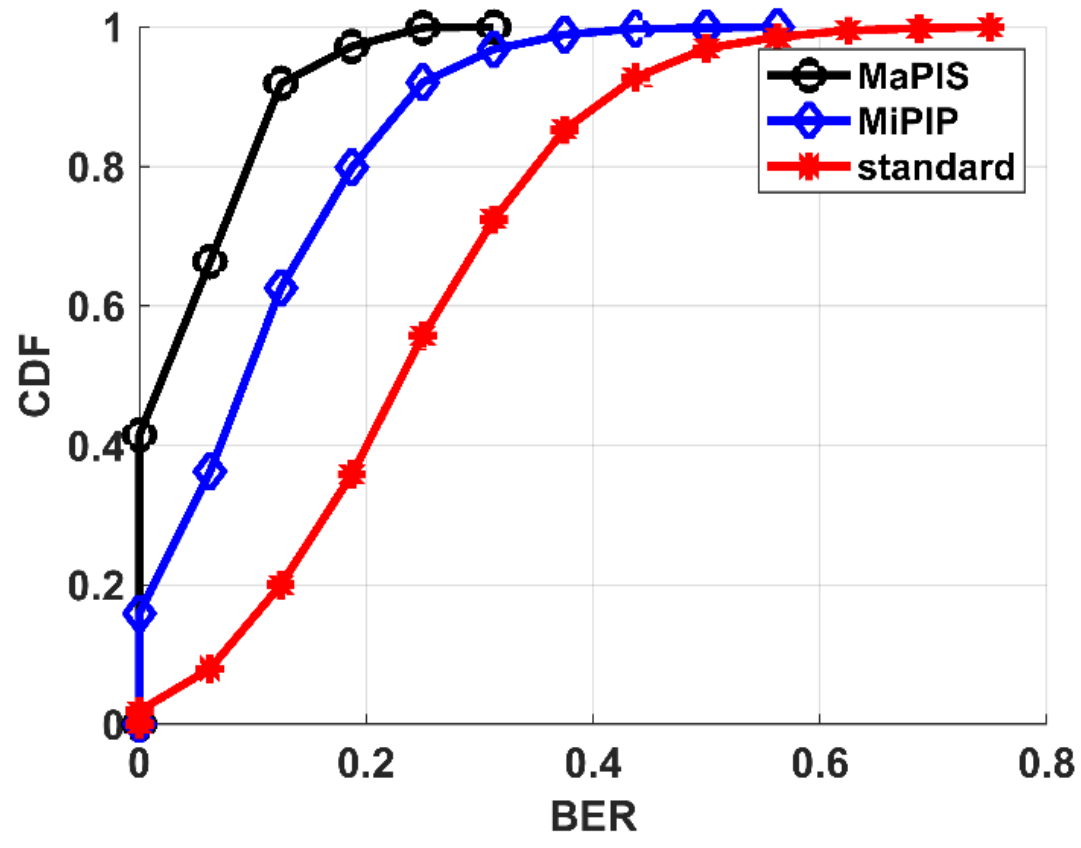

*(b)*

**FIGURE 21. Comparison of (*a*) CDFs of SER and, (*b*) CDFs of BER for the proposed and standard methods in a high CCI case.**

### 6) FULLY-BLOCKED OWN-CHANNEL CASE ($h_{ii} = 0$)

The average SER and average BER results are shown in FIGURE 22*(a)* and *(b)* respectively. As seen from the figure, the best performance is achieved by the proposed MaPRIS by far. The standard method and the proposed MiPIP give same results as expected because own-signal is zero ($h_{ii} = 0$).

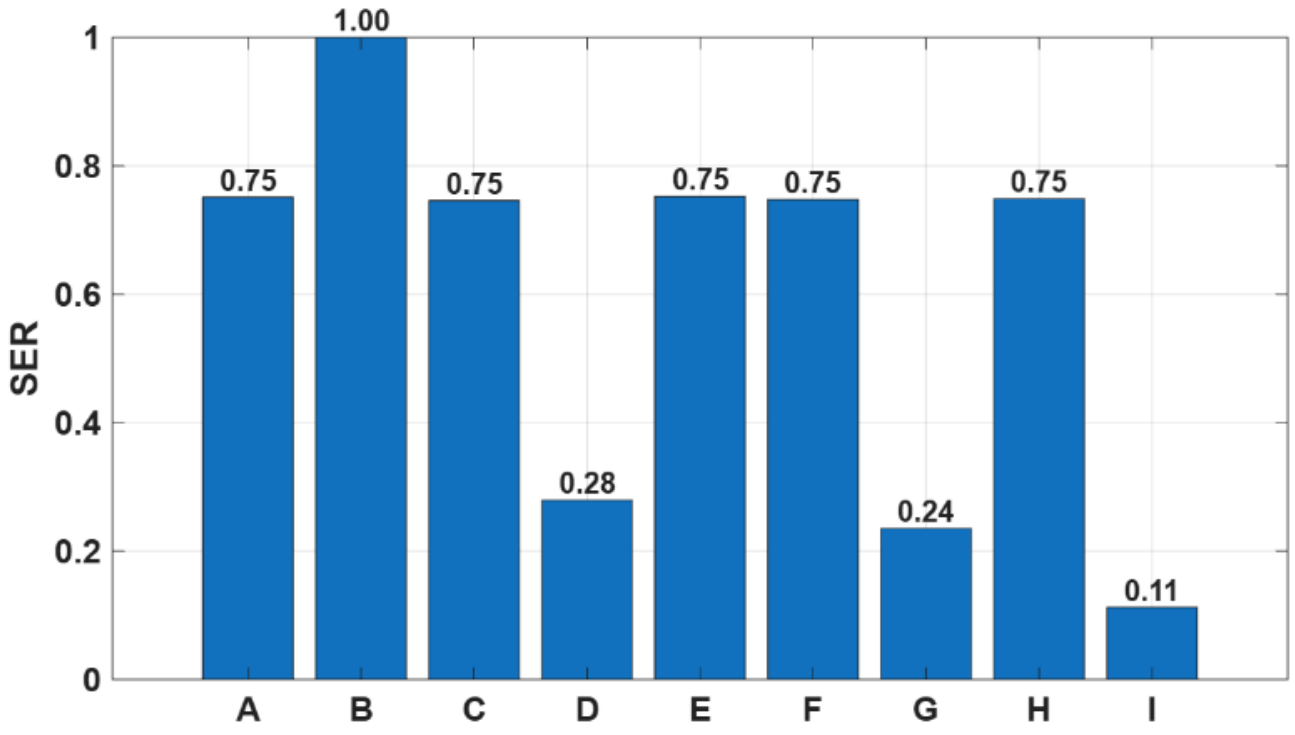

*(a)*

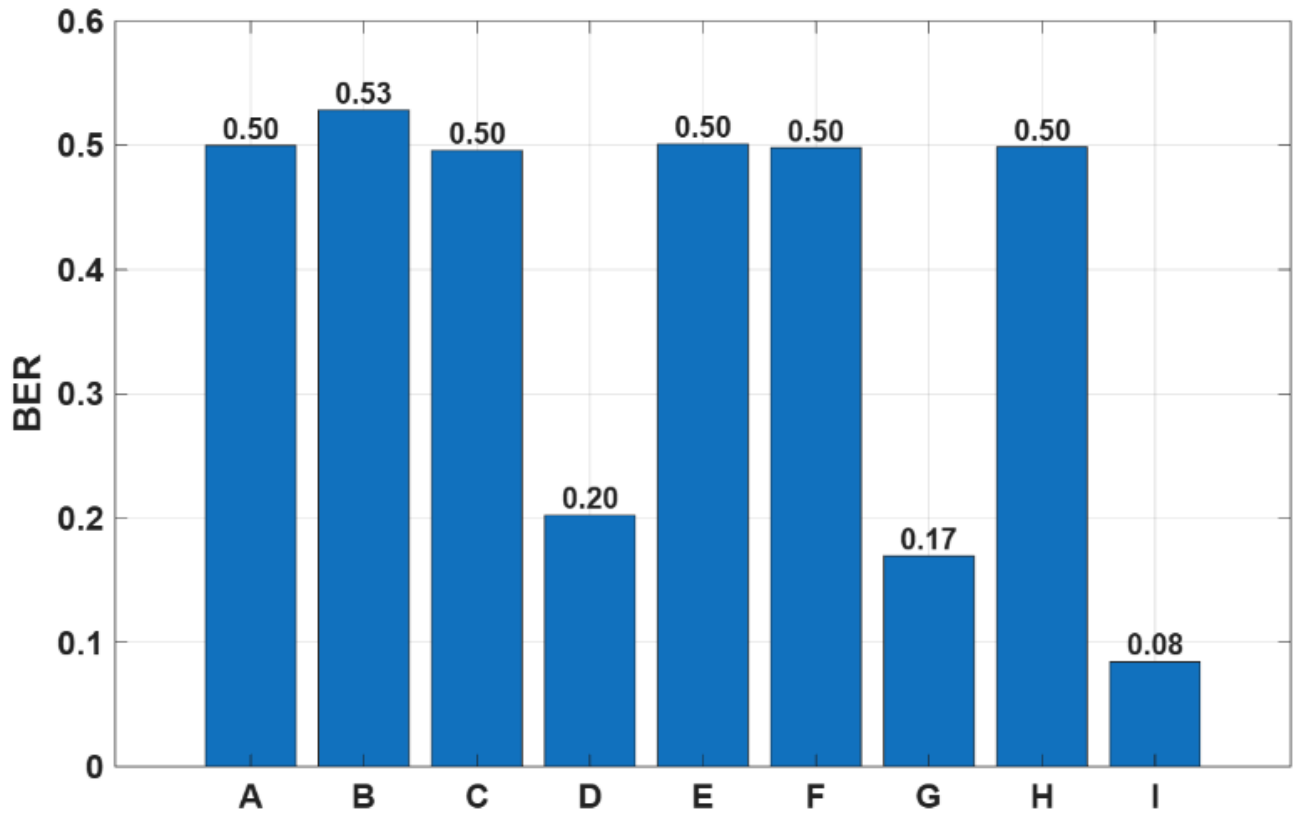

*(b)*

**FIGURE 22. (*a*) Overall average SER results, and (*b*) BER results for the $h_{ii} = 0$ case.**

CDF of SER and CDF of BER of the traditional system in FIGURE 2 and the proposed methods MaPRIS and MiPIP are presented in FIGURE 23.(*a*) and FIGURE 23.(*b*), respectively (the fully-blocked own-channel case). The MaPRIS gives superior performance as compared to all other methods. The 2nd best performance is obtained by the CHNN and the SIGA when all desired channel links are *simultaneously completely blocked* (i.e., $h_{ii} = 0$).

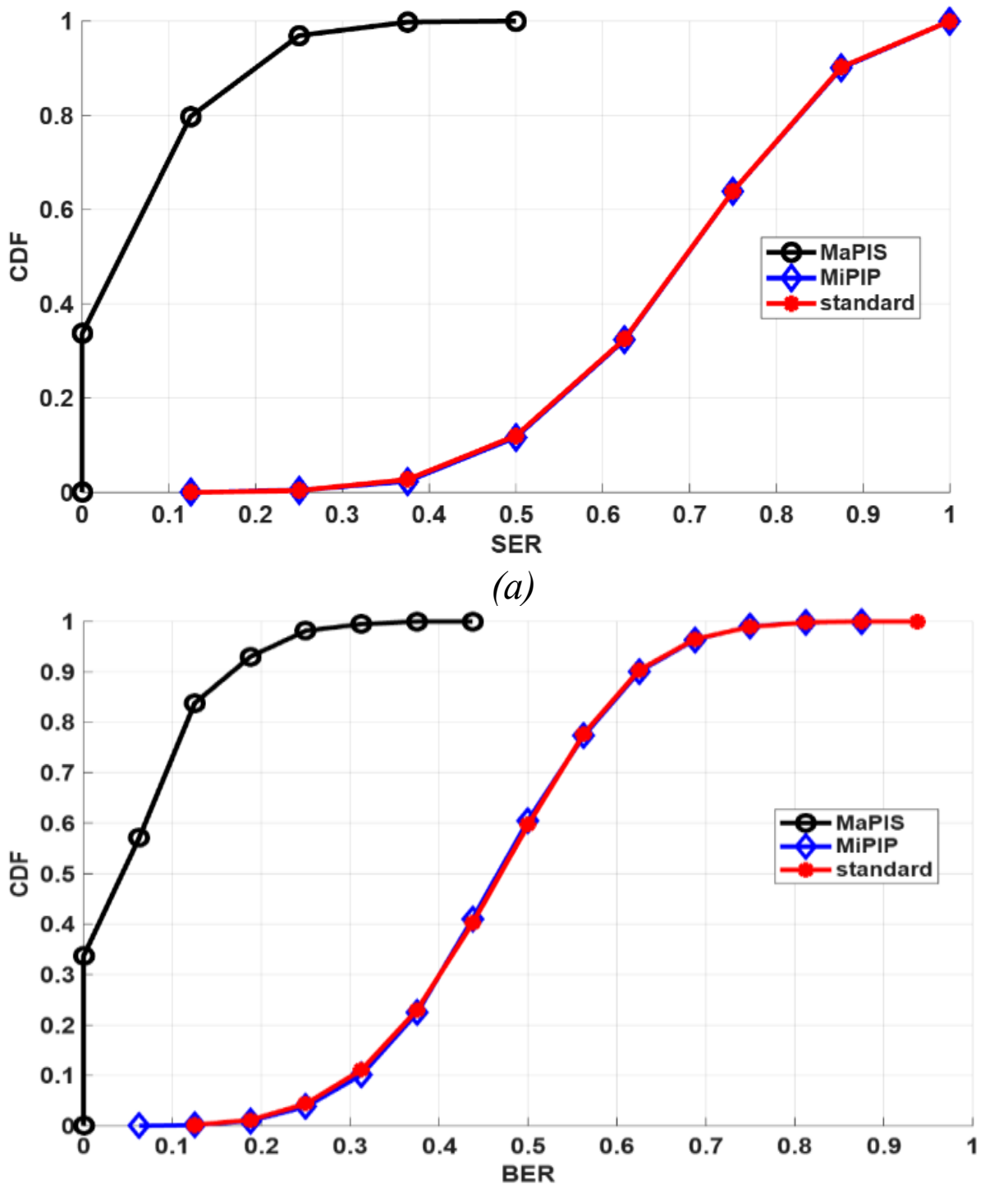

*(a)*

*(b)*

**FIGURE 23. Comparison of (*a*) CDFs of SER and (*b*) CDFs of BER for the proposed and standard methods in the fully-blocked own-channel case.**

## VII. CONCLUSIONS

The co-channel interference (CCI) continues to be one of the primary obstacles to realizing the high data rate objectives of current and future wireless communication systems which are inevitably to become CCI-limited due to the aggressive spectrum reuses in dense deployments. The growing demand for advanced CCI management drives researchers to explore innovative techniques that leverage cutting-edge signal processing and optimization frameworks.

In this paper we address the CCI challenge in the CCI-limited wireless scenarios. Being inspired by the concept called shadow-cuts in [2], we reformulate/turn classical co-channel wireless communication systems into a novel phase-optimized framework, where each transmitter and receiver is associated with a discrete phase state. This leads to a novel optimization problem, which we called "Game of Phases: Surrogate Symbol Optimization": Given a channel gain matrix and a discrete phase set, we seek to determine the optimal surrogate symbols (i.e. “proxies”) that minimize average BER of overall wireless system.

As an alternative to the traditional wireless communication systems, we, in this paper, present a novel “surrogate symbol” transmission framework wherein a proxy symbol is selected and transmitted instead of the intended desired symbol.

For solving the Surrogate Symbol Optimization problem, we present two novel phase optimization algorithms for finding the optimum surrogate symbols for a given channel matrix: Minimum Phased-Interference Power algorithm (MiPIP) and Maximum Phased-Interference Support (MaPRIS). We show that the MiPIP reduces the BER by minimizing the variance of the phased interference.

The proposed method MaPRIS converts the “hostile” CCI into a “friendly signal” (named as “interference support” in our paper), which yields the following counter-intuitive result: The higher the co-channel interference variance is, the higher the SINR is, and thus the higher the data rate is. *The MaPRIS achieves this by minimizing the angle between the rotated surrogate symbol vector and the surrogate interference vector at the centralized unit. We call this angle as* “effective angle”. *This very angle turns out to be equal to the angle between the desired symbol and the phased and rotated interference vector at the receiver. This is the key merit of the MaPRIS which converts the hostile co-channel interference into a “friendly signal” which is close-to the desired symbol.*

Even if the desired signal is totally blocked for any rx (i.e., even if the own channel coefficient is zero), there is still a (high) probability that the desired symbol be recovered. This is because the MaPRIS turn the CCI angle into “almost” the desired symbol angle for each and every rx.

Extensive simulation results confirm the findings and show that the proposed MaPRIS and MiPIP give remarkably superior BER performance as compared to the conventional transmission schemes and the other phase selection algorithms when $\boldsymbol{h}_{ii} \neq 0$. For the fully-blocked ($\boldsymbol{h}_{ii} = 0$) case, the best results are obtained by the MaPRIS by far and the 2nd best results are obtained by the CHNN and SIGA [2] based phase optimization algorithms.

The system level simulation results presented in Section-V serve as an upper bound to the proposed surrogate-symbol based wireless system in **FIGURE 4** due to assumption 4 (i.e., each tx sends the rx-phase index to the corresponding rx from a separate and CCI-free channel).

In this paper, we focus on single-antenna case. Investigating the MIMO case is an interesting research direction in the future. We reiterate that because the proposed algorithms act on the baseband *effective* channel matrix **H** alone, the MaPRIS and MiPIP gains are realized on top of, not instead of, whatever power-control, beamforming, CoMP, or interference-cancellation techniques already shape that effective channel; the proposed framework and these other techniques are therefore complementary rather than competing.

More broadly, this paper shows that CCI, rather than being merely tolerated or suppressed, can be reshaped by phase optimization alone into a contribution to the desired symbol and, in the limiting case, into a proxy for it even when the desired channel is completely blocked. The analysis assumes perfect CSI, a centralized phase-assignment unit and a CCI-free control channel. As is standard for establishing performance limits, the reported results are therefore an upper bound on the performance of practical systems (Section III-C). Natural directions for future work are: (i) performance under imperfect CSI and channel-estimation error, including the sensitivity of the interference-support mechanism to phase-estimation error; (ii) decentralized phase assignment with reduced signaling; (iii) alternative optimization objectives such as max-min (worst-user) rate, which trade average performance for fairness; and (iv) the computational and signaling scalability of the centralized phase assignment as $N$ grows, including measured running time for large-scale networks.

## APPENDIX A

### *HOW TO ESTIMATE THE N×N CO-CHANNEL MATRIX H IN CELLULAR RADIO SYSTEMS*

This appendix outlines key methodologies for acquiring, representing, and exchanging channel state information (CSI) needed to form the network wide co-channel matrix **H** across $N$ transmit points (e.g., base stations) and $N$ receive points (e.g., user terminals).

#### 1) PILOT-BASED (CLASSICAL) CHANNEL ESTIMATION

Transmitters send mutually orthogonal known pilot signals. Then the received pilot signals $\boldsymbol{Y} = \boldsymbol{H}\boldsymbol{\Pi} + \boldsymbol{N}$, where $\boldsymbol{N}$ is background noise. Since $\boldsymbol{\Pi} = \mathbf{I}$, then $\boldsymbol{Y} \approx \boldsymbol{H}$. This is done during a training phase (see e.g. [12]). In cellular systems, Base Stations (BSs) coordinate this process, with User Equipment (UE) feeding back decomposed Channel State Information (CSI) (e.g., codebook indices) to the BS. The original CSI is recovered at the BS because the codebooks are already known at the BS [16]. Because BSs in the neighborhood are connected to the same central Controller, the channel matrix **H** is available at the central Controller coordinating BSs.

#### 2) CHANNEL RECIPROCITY IN TDD SYSTEMS

In Time Division Duplex (TDD) systems, channel reciprocity allows BSs to infer downlink channels from uplink pilots, especially in massive MIMO systems.

#### 3) DATA DRIVEN / DEEP LEARNING-BASED CHANNEL PREDICTION

Deep learning (DL) offers a breakthrough in channel estimation [14]. DL models predict **H** using inputs like previous channel realizations, environmental data, signal strength (RSSI), CSI, Angle of Arrival (AoA) information, etc.. Common DL architectures include RNNs/LSTMs and CNNs [13], [14]. For further references, see, e.g., [13].

#### 4) COMPRESSED SENSING / SPARSE ESTIMATION

Sparse channels (e.g., mmWave and RIS channels) enable compressed sensing with fewer pilots, as demonstrated for millimeter-wave massive MIMO systems using separable compressive sensing techniques [15]. For example, very recently, a deep learning-based quantized trainable compressed sensing CSI feedback method is proposed [16]. Their simulations demonstrate that their method outperforms the existing methods in 3GPP UMi and UMa channels in reducing overhead while maintaining accuracy.

#### 5) TRANSFER AND FEDERATED LEARNING

Transfer learning adapts pre-trained CSI models to new sites. In other words, we may use an CSI model trained in one environment to fine-tune in another environment [17]. Federated schemes combine distributed updates without raw CSI sharing. For further references, see, e.g. [17].

Some representative patents addressing efficient CSI exchange and feedback mechanisms in wireless networks are summarized as follows:

[18] proposes methods for efficient CSI exchange between BSs for CoMP transmission and reception.

[19] discloses a method for efficient CSI transmission using hybrid CSI reference signals for enhanced feedback accuracy.

[22] describes a system for configuring CSI reports to match CoMP transmission hypotheses, improving spectral efficiency.

[25] focuses on reducing CSI feedback overhead in CoMP environments by exploiting correlations between multiple CSI reports through compression.

[24] details a method for UEs to selectively report single or multiple CSIs to reduce feedback overhead.

[23] describes methods and apparatus for CoMP CSI feedback, enabling UEs to perform separate interference measurements and transmit corresponding reports under various channel and interference assumptions.

## APPENDIX B

In this Appendix, we present the receiver-specific effective angles $\left|\Xi_i^{eff}\right|$ in the initial step and final step of the MaPRIS with $J_{MaPIS}$ in (34) for the same channel realization in FIGURE 12 in Section V.A.

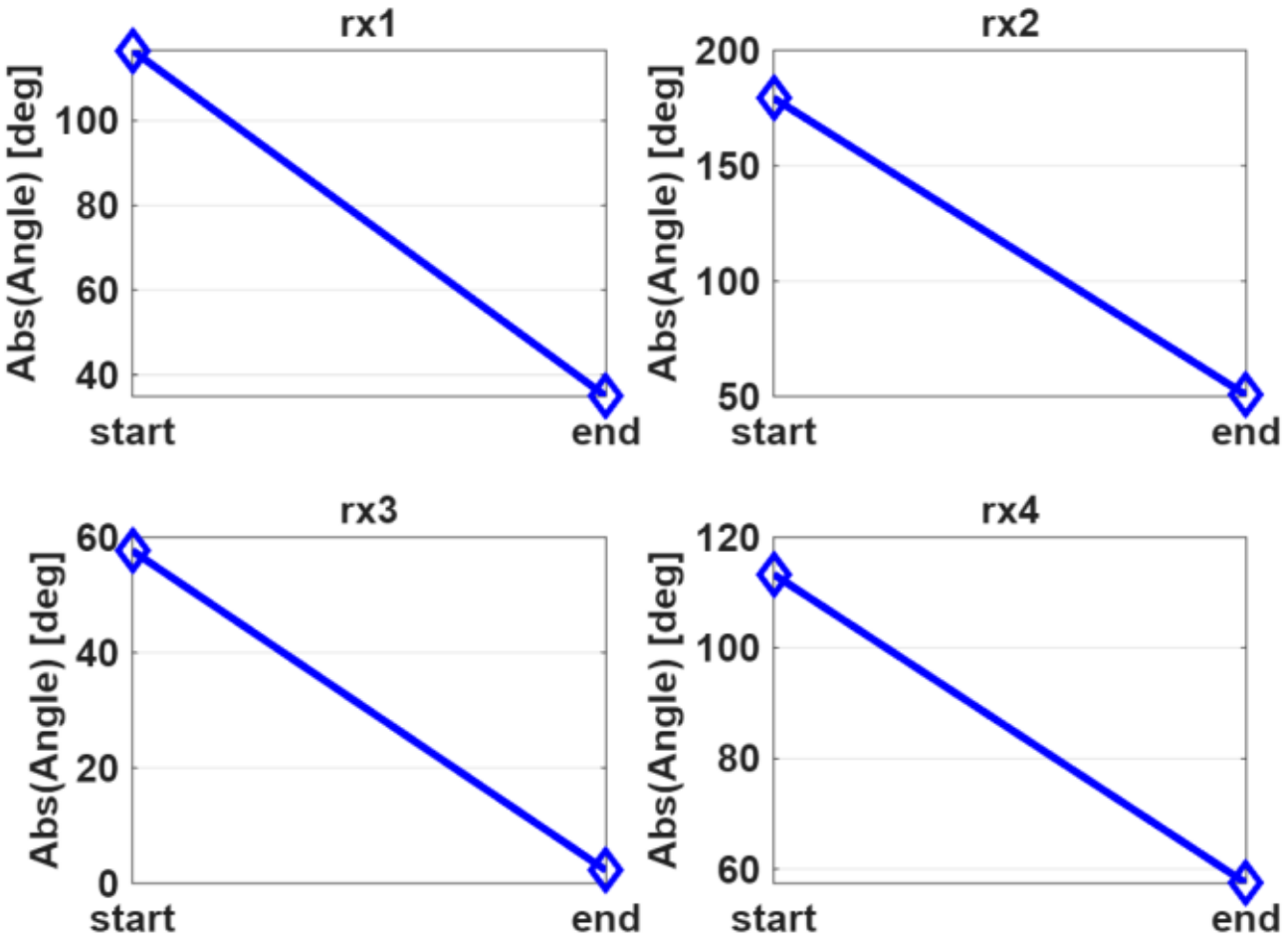

*(a)* Receivers from 1 to 4

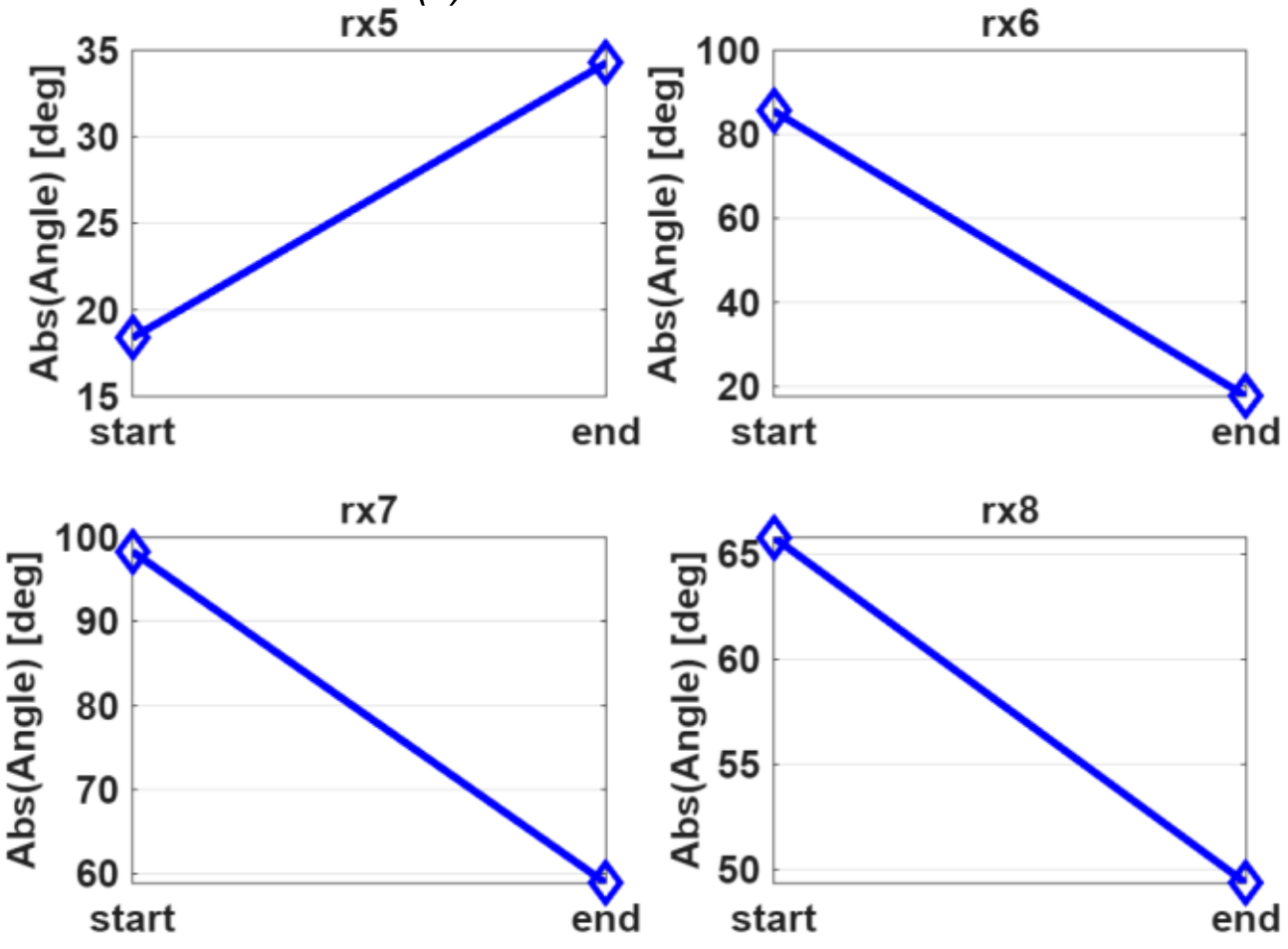

(b) Receivers from 5 to 8

**FIGURE 24. The receiver-specific effective angles $|\Xi_i^{eff}|$ in the initial step and final step of the MaPRIS for the same channel realization in Section V.A.1.**

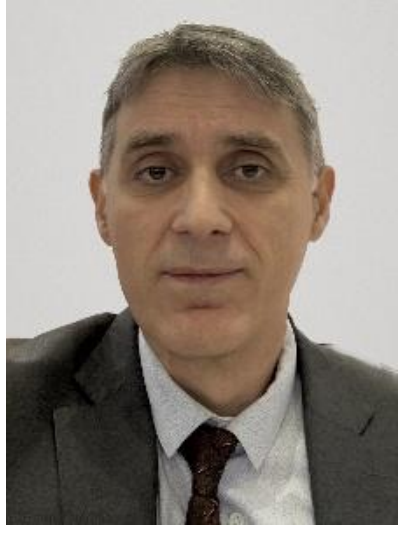

**Zekeriya UYKAN** (M’94–SM’10) received the B.S. and M.S. degrees (Hons.) from Istanbul Technical University (ITU), Istanbul, Turkey, in 1993 and 1996, respectively, and the Lic.Tech. (Hons.) and Dr.Tech. degrees from the Helsinki University of Technology (HUT, now Aalto University), Espoo, Finland, in 1998 and 2001, respectively, all in electrical engineering.

He was with the NOKIA Research Center (now Nokia-Bell Labs), Radio Comm. Laboratory, Helsinki, Finland, from July 2000 to April 2007, and Nokia Siemens Networks (now Nokia Solutions and Networks), Espoo, from April 2007 to December 2010. He was a Visiting Scholar with Stanford University, Stanford, CA, USA, from January 2002 to December 2002; a Visiting Scientist with Harvard University, Cambridge, MA, USA, from September 2008 to June 2009. He was with Dogus University, Istanbul, from September 2009 to September 2015, and has been with the American University of the Middle East, Egaila, Kuwait, since September 2014 where he is currently a Full Professor. His research interests include artificial intelligence, fuzzy systems and wireless communications.